\documentclass[twocolumn]{aastex631}

\newcommand\CII{[\ion{C}{2}]}
\newcommand\kms{km s$^{-1}$}

\begin{document}

\title{ALMA and JWST Imaging of $z\ >\ 6$ Quasars: No Spatial Position Offset Observed Between Quasars and Their Host Galaxies}

\author[0009-0002-4145-1011]{Aurora Wilde}
\affiliation{National Radio Astronomy Observatory, 520 Edgemont Rd, Charlottesville, VA 22903 USA}
\affiliation{Steward Observatory, University of Arizona, 933 N Cherry Ave, Tucson, AZ 85719 USA}

\author[0000-0002-9838-8191]{Marcel Neeleman}
\affiliation{National Radio Astronomy Observatory, 520 Edgemont Rd, Charlottesville, VA 22903 USA}

\author[0000-0001-5492-4522]{Romain A. Meyer}
\affiliation{Department of Astronomy, University of Geneva, Chemin Pegasi 51, 1290 Versoix, Switzerland}

\author[0000-0002-2662-8803]{Roberto Decarli}
\affiliation{INAF – Osservatorio di Astrofisica e Scienza dello Spazio
di Bologna, via Gobetti 93/3, I-40129, Bologna, Italy}

\author[0000-0003-4793-7880]{Fabian Walter}
\affiliation{Max Planck Institut f$\ddot{u}$r Astronomie, K$\ddot{o}$nigstuhl 17, D-69117 Heidelberg, Germany}

\author[0000-0003-1625-8009]{Brenda L. Frye}
\affiliation{Steward Observatory, University of Arizona, 933 N Cherry Ave, Tucson, AZ 85719 USA}

\author[0000-0003-3310-0131]{Xiaohui Fan}
\affiliation{Steward Observatory, University of Arizona, 933 N Cherry Ave, Tucson, AZ 85719 USA}

\begin{abstract}
We present a study determining the spatial offset between the position of the supermassive black hole (as traced through their broad line regions) and the host galaxy in six $z > 6$ quasars. We determined the host galaxy's position from $\lesssim$$0\farcs10$ ($\lesssim$ 600 pc) resolution Atacama Large Millimeter/sub-millimeter Array (ALMA) \CII\ 158 $\mu m$ and corresponding dust continuum imaging. We determined the quasar's position from $\lesssim$ 400 pc resolution James Webb Space Telescope Near-Infrared Camera (JWST NIRCam) imaging. We estimated the observational uncertainties on the quasar's position using astrometric data from the Global Astrometric Interferometer for Astrophysics (GAIA) of field stars within the NIRCam images. We find that all six quasars are found within the central $\sim 400$ pc of their host galaxy dust continuum and \CII\ emission. Apparent offsets seen in rest-frame optical JWST observations are not detected in our ALMA data, suggesting they likely result from dust obscuration rather than a true physical separation between the SMBH and its host galaxy. Kinematic modeling of these data further reveals that none of the galaxies show evidence for recent merger activity, and most of the galaxies can be accurately modeled using a simple disk model. The lack of an offset supports theoretical models that predict that positional offset within these galaxies are either short-lived or intrinsically rare.

\end{abstract}


\section{Introduction} \label{sec:intro}

Quasi-stellar objects (QSOs) or quasars\footnote{In this paper, we will use QSO and quasar interchangeably. Although most objects in this manuscript do not show significant radio emission and therefore technically  do not satisfy the definition of a quasar, it nevertheless has been common practice in the literature to label them as quasars. Furthermore, we will use the term quasar only for the AGN region, and reserve the term quasar host galaxy for the galaxy hosting the quasar.} are active galactic nuclei (AGNs) observed off-axis from a possible radio jet, with an unobscured view of both the narrow- and broad-line regions. These AGN are powered by accreting gas onto a supermassive black hole (SMBH). Quasars can be observed out to great distances \citep[redshifts greater than 7.5;][]{Banados_2017, Wang_2021}, and have been shown to have properties that can be intricately linked with the properties of the galaxies that host them. In particular, observations of quasars have indicated a correlation between SMBH masses and their host galaxies' bulge mass, luminosity, and velocity dispersion \citep[][for a comprehensive review]{Beckman_1993, Ferrarese_Merritt_2000, Gebhardt_2000, Merritt_2001, Kormendy_Gebhardt_2001, Marconi_Hunt_2003, McConnell_2013, Kormendy_2020}. While the nature of a coevolutionary relationship between SMBHs and their host galaxies (if one exists at all) is still not exactly understood, studying luminous quasars at high redshift ($z \gtrsim 6$) opens up an avenue to explore galaxy formation and the galaxy -- black hole connection in the universe's first billion years.

To study the properties of the SMBHs that power $z \gtrsim 6$ quasars, most studies rely on the bright, rest-frame ultraviolet and optical emission that arises from the quasar. In particular, black hole masses are estimated primarily through broad emission lines such as \ion{Mg}{2} and H$\beta$, and estimators calibrated at lower redshift \citep{Vestergaard_2009,  Shen_2011, Mazzucchelli_2017, Farina_2022, Yang_2023}. To study the host galaxies of these quasars, the Atacama Large Millimeter/sub-millimeter Array (ALMA) has transformed our understanding of these galaxies by spatially resolving them at unprecedented resolution \citep[e.g.,][]{Decarli_2018, Venemans_2020, Fan_2023}. Unlike the optical/near-infrared regime, where the quasar outshines its host galaxy by orders of magnitude, at the frequencies ALMA observes, the galaxy emission dominates over the quasar emission. Using fine-structure line transitions, such as the \CII-158 $\mu$m (\CII) emission line, and dust continuum emission, the host galaxies of these quasars are being mapped in exquisite detail \citep{Venemans_2019, Yue_2021, Walter_2022, Shao2022, Tripodi_2023, Neeleman_2023, Meyer_2025}. These observations have revealed that $z \gtrsim 6$ quasar host galaxies have relatively compact interstellar medium (ISM) and a high gas mass fraction \citep{ Walter_2009, Pensabene_2020, Neeleman_2021}. This is in stark contrast with the galaxies in the nearby universe that host such massive SMBHs \citep{Pensabene_2020, Almeida_2022}, and suggests that the host galaxies of $z \gtrsim 6$ quasars are still forming and acquiring mass. Comparison between the dynamical mass of the host galaxy and the SMBH mass further reveals that black hole mass estimates are generally much larger than what is expected based on the correlation between these two masses in the local universe \citep[e.g.][]{Pensabene_2020, Neeleman_2021}. This again suggests that the host galaxies of $z \gtrsim 6$ quasars are actively accreting material, and are still in the process of evolving into more massive galaxies. It should be noted that numerical simulations predict that to reproduce these observed large black hole masses so early in the universe, they either started with large black hole seed masses and/or had accretion rates at or above the Eddington limit \citep{ Farina_2022, Volonteri_2023, Schneider_2023, Bennett_2023}. Such extreme formation mechanisms occur in only the densest regions of the universe, and luminous quasars are therefore not necessarily representative of the typical AGNs at these redshifts. Nevertheless, understanding how the most massive objects formed provides crucial information on the extreme limits of the processes that govern early galaxy formation.

Even accounting for large black hole seed masses, mergers with satellite galaxies are integral to the mass assembly of the very massive galaxies that host these luminous sources \citep{Haehnelt_2002}. This process of merging galaxies facilitates the formation of many black hole binaries. Simulations suggest that merging black holes may experience a recoil effect, resulting in a position or velocity offset between the emergent black hole and the galaxy's mass center \citep{Haehnelt_2002, Campanelli_2007, Blecha_2016, Dong-Paez_2025}. Both spatial offsets and recoil velocities have been observed in nearby galaxies \citep{Civano_2012, Chiaberge_2017, Chiaberge_2018}. The maximum expected recoil velocities are on the order of $\sim10^3\ $ \kms, with offsets being sustained for up to a Gyr \citep{Campanelli_2007, Blecha_2016}. Two of the most important factors in producing an offset after a merger are the mass ratio of the black holes and the alignment/anti-alignment of their spins \citep{Campanelli_2007, Volonteri_2010, Dotti_2012, Blecha_2016, Dong-Paez_2025}. Significant recoil events require a nearly equal black hole mass ratio, and are expected to be less common for the most massive SMBHs due to the scarcity of similarly massive counterparts \citep{Dotti_2012, Blecha_2016, Dunn_2020}. In addition, recoil velocities are much larger in systems where black hole spins are significantly misaligned \citep{Campanelli_2007, Dotti_2012, Blecha_2016}. In wet mergers (i.e., a merger of a gas-rich galaxy), which are expected to be common at z $\gtrsim$ 6, dynamical friction with surrounding gas may facilitate alignment of black hole spins prior to merging \citep{Volonteri_2010, Dunn_2020}. Wet mergers also funnel cold gas to the center of the galaxy, increasing the escape velocity for SMBHs in the center and further suppressing recoil events \citep{Volonteri_2010, Blecha_2016, Dunn_2020}. Large recoil events reduce black hole growth both from accretion and future mergers, which would make a recoil history unlikely for very massive SMBHs at high redshift \citep{Dunn_2020, Dong-Paez_2025}. Although all these points disfavor significant black hole recoil events in $z \gtrsim 6$ quasars, with the advent of ALMA and the James Webb Space Telescope (JWST) this hypothesis can now be tested.

In the local universe, galaxies with SMBHs that have roughly the same mass as the luminous quasars studied at $z \gtrsim 6$ are all found within the central regions of predominantly massive elliptical galaxies \citep[e.g.,][]{Klutse_2024}. This picture changes significantly at high redshift, where quasars are found in a range of different galaxies, including disks, dispersion-dominated systems, and complex mergers \citep{Neeleman_2021}. Since galaxy mergers are expected to precipitate black hole mergers, quasars with positional offsets may also display kinematic signatures indicative of recent mergers, such as disturbed velocity fields or tidal features, unless these features dissipate rapidly \citep{Hopkins_2008}. However, ALMA observations of a sample of 27 $z \gtrsim 6$ quasar host galaxies at a resolution of $\sim0\farcs25$ revealed no significant offset between the optical position of the quasar (from ground-based facilities) and the center of the galaxy \citep{Venemans_2020} independent of the morphology of the galaxy. However, the ability to measure offsets in this study is hampered by the resolution of the data. Intriguingly, recent higher resolution observations at $z \gtrsim 6$ have shown the first possible offsets between the center of the galaxy and the quasar position \citep{Meyer_2023, Ding_2023, Ubler_2024}. In particular, \citet{Meyer_2023} imaged QSO~J0109–3047 at a resolution of 300\,pc with ALMA, and found that the black hole appeared displaced with respect to both the \CII\ and continuum emission. These high resolution ALMA observations are crucial in order to fully characterize the galaxy morphology.

In this paper, we expand on the results in \citet{Meyer_2023}, and analyze archival ALMA observations of six quasars at a minimum resolution of $0\farcs1$ in order to investigate the dynamics of these systems and search for evidence of mergers. These systems have also been observed with JWST, and we will use the images obtained with JWST to compare the near-infrared position of the quasar with the ALMA observations to search for possible spatial offsets between the galaxy and the quasar, which could be a possible sign post of a black hole recoil. In addition, we will perform kinematic modeling of each host galaxy to detect any other features which may indicate a merger. The paper is structured as follows. In Section \ref{sec:method} we detail the collection of quasars, their data sets, and the data reduction process. The resulting analysis of the continuum and \CII\ data are found in Section \ref{sec:analysis}. This section also contains the astrometric comparison to JWST images and the modeling of the kinematics of the galaxy. Discussion of the results can be found in Section \ref{sec:discussion}. Conclusions are in Section \ref{sec:conc}. Throughout this paper we use a standard flat $\Lambda$ cold dark matter cosmology with $H_0 = 70$ \kms Mpc$^{-1}$ and $\Omega_M = 0.3$. For this cosmology, 1$''$ corresponds to 5.7 kpc at $z = 6$.

\section{Sample Selection \& Data Reduction} \label{sec:method}
\subsection{ALMA data} \label{subsec:alma}

\begin{deluxetable*}{ccccccccccc}
\tabletypesize{\scriptsize} 
\tablecaption{ALMA Data Set of High Z Quasars: Observation Information \label{tab:alma obs}}
\tablehead{
\colhead{Quasar} & \colhead{Project ID}& \colhead{Cycle}& \colhead{RA}& \colhead{Dec}& \colhead{Int. Time}& \colhead{Cont. Res.}& \colhead{Cont. Sensitivity}& \colhead{\CII\ Sensitivity}& \colhead{$L_{\rm{bol}}$}\\
\colhead{} & \colhead{}& \colhead{}& \colhead{(h:m:s)}& \colhead{(d:m:s)} & \colhead{($\rm sec$)}& \colhead{($\rm as$)}& \colhead{($\rm mJy/beam$)}& \colhead{($\rm mJy/beam$)}& \colhead{$10^{46} ~\rm erg~s^{-1}$}\\
}
\startdata 
{          }& 2019.1.00672.S& 7& 20:54:06.48& $-$00:05:14.800& 2377&  \\ 
{J2054$-$0005}& 2019.1.00672.S& 7& 20:54:06.48& $-$00:05:14.800& 1814& 0.073& 0.010& 0.12& 12.37  \\ 
{          }& 2018.1.00908.S& 6& 20:54:06.42& $-$00:05:14.800& 5141&   \\   
\hline
{J0923$+$0402}& 2021.1.00934.S& 8& 09:23:47.117& $+$04:02:54.580& 6985& 0.095& 0.0086& 0.13& 21.70\\ 
{          }& 2018.1.01188.S& 6& 09:23:47.117& $+$04:02:54.580& 816&  \\
\hline
{J2002$-$3013}& 2021.1.00934.S&  8&  20:02:41.594& $-$30:13:21.690& 6894& 0.10& 0.014& 0.15& 15.4\\ 
{          }& 2019.1.01025.S&  7& 20:02:41.594& $-$30:13:21.690& 756& \\ 
\hline
{J2102$-$1458}& 2021.1.00934.S&  8& 21:02:19.230& $-$14:58:53.860& 8013& 0.12& 0.0099& 0.15& 6.0\\  
\hline
{J0244$-$5008}& 2021.1.00934.S&  9&  02:44:01.020& $-$50:08:53.700& 4971& 0.089& 0.0094& 0.14& 14.4\\  
\hline
{J0109$-$3047}& 2015.1.00399.S& 2& 01:09:53.130& $-$30:47:26.300& 1966& 0.044& 0.0063& 0.07& 7.69 $\pm$ 0.20&\\
{}& 2021.1.00800.S& 8& 01:09:53.130& $-$30:47:26.319& 26291& \\
\enddata
\tablecomments{This is a complete list of ALMA data sets used for each source. All sets use ALMA band 6, which includes the \CII\ line at these redshifts. The last four columns, which list only one value per source even for sources with multiple observation IDs, are measurements of the combined observations. Bolometric Luminosities for each source are sourced from previous publications (J2054$-$0005 from \citet{Farina_2022}; J0923$+$0402, J2002$-$3013, and J2102$-$1458 from \citet{Yang_2021}; and J0244$-$5008 from \citet{Reed_2019}). J0109$-$3047 values are from analysis done in \cite{Meyer_2023}. }
\end{deluxetable*}

We selected quasars from the ALMA archive with band 6 observations targeting redshifted \CII\ emission and angular resolution better than $0\farcs1$. Thirteen quasars met these criteria and had available JWST NIRCam stage 3 imaging. We exclude P036+03 from our sample due to a nearby galaxy in the NIRCam image whose overlapping emission prevents a reliable measurement of the quasar’s position. We included all available ALMA observations for the remaining sources in our data reduction, including supplemental lower-resolution data. Five sources had combined integration times exceeding 2 hours. We selected four of these high-integration sources for analysis, excluding the lensed quasar J0439$+$1634. We also included one lower-integration source (J0244$-$5008) to test whether shorter observations are sufficiently accurate for our analysis. Additionally, we include the source J0109$-$3047 (which was discussed in \cite{Meyer_2023}) in our sample. This source is excluded from the imaging step of our analysis, as that was performed in \cite{Meyer_2023}. We note that the reduction and imaging of this source was identical to the other sources in the sample. Each observation's project ID, cycle, pointing, and integration time are listed in Table \ref{tab:alma obs}. 

All ALMA data were processed using the ALMA pipeline \citep{Hunter2023}, which is part of the Common Astronomy Software Application \citep[CASA;][]{CASA2022} package. The CASA version used for each data set varied depending on the observation cycle, where we used the version of the pipeline that was initially used to generate the data sets. 

Imaging was performed with the task \textit{tclean} in CASA. Continuum images were cleaned using Briggs weighting with a robust parameter of 0.5, down to twice the root-mean-square (rms) noise near the quasar center. A pixel size of $0\farcs01$ was used, which insured sufficient sampling of the synthesized beam. All images were cleaned using multiscale cleaning. The synthesized clean beam and sensitivity are listed in Table \ref{tab:alma obs}. The \CII\ emission line was imaged with 30 \kms\ channel width, using the same Briggs robust parameter and pixel scale, and similarly cleaned down to twice the rms noise of the channel. A primary beam correction was applied, although we note that the effect was negligible, because our sources are close to the phase center of the ALMA observations.

\subsection{JWST data} \label{sec:jwst}

\begin{deluxetable*}{cccccccccccccc}
\tabletypesize{\scriptsize}
\tablecaption{JWST NIRCam Quasar Data Set Properties\label{tab:jwst}}
\tablehead{
\colhead{Quasar} & \colhead{R.A.}& \colhead{$\sigma_{\rm R.A.}$}& \colhead{Decl.}&  \colhead{$\sigma_{\rm Decl.}$}& \colhead{$\Delta_{\rm cont}$}& \colhead{\textbf{$\sigma_{\Delta \rm cont}$}}& \colhead{$\Delta_{\text{\CII}}$}& \colhead{$\sigma_{\Delta \text{\CII}}$}& \colhead{Program Number}& \colhead{Filter}& \colhead{Exposure Time} \\
\colhead{   }& \colhead{(h:m:s)}& \colhead{\textbf{(mas)}}& \colhead{(d:m:s)}& \colhead{\textbf{(mas)}}& \colhead{(mas)}& \colhead{(mas)}& \colhead{(mas)}& \colhead{(mas)}& \colhead{ }& \colhead{ }& \colhead{(sec)}  \\
}
\startdata 
J2054$-$0005& 20:54:06.50& 50& $-$00:05:14.40& 60& 70& 55& 60& 56& 1813& F356W;CLEAR& 4982 \\
J0923$+$0402& 09:23:47.12& 60& $+$04:02:54.36& 50& 40& 58& 150& 60& 2078& F356W:CLEAR& 1417\\
J2002$-$3013& 20:02:41.59& 50& $-$30:13:21.56& 60& 30& 55& 30& 59&  2078& F115W;CLEAR& 1417\\
J2102$-$1458& 21:02:19.22& 60& $-$14:58:54.02&  50& 10& 55& 20& 59& 2078& F356W;CLEAR& 1417\\
J0244$-$5008& 02:44:01.04& 150& $-$50:08:53.75& 170& 50& 160& 50& 161& 2078& F115W;CLEAR& 1417\\
J0109$-$3047& 01:09:53.13& 130& $-$30:47:26.34& 130& 60& 130& 60& 133& 2078& F356W;CLEAR& 1417\\
\enddata
\tablecomments{
Right Ascension and Declination listed are the center of quasar emission, as described in Section \ref{subsec:astrometry}. $\sigma_{\rm R.A.}$ and $\sigma_{Decl.}$ are the standard deviation of the offsets between background stars' GAIA coordinates and JWST NIRCAM position. Stage 3 data products were used for all sources. $\Delta_{\rm cont}$ and $\Delta_{\text{\CII}}$ are the offset between the listed SMBH coordinates and the peak of continuum and \CII\ emission of their host galaxies, respectively. $\sigma_{\Delta \rm cont}$ and $\sigma_{\Delta \text{\CII}}$ are their respective errors, computed by adding the average of $\sigma_{\rm R.A.}$ and $\sigma_{\rm Decl.}$ (JWST position uncertainties) to $\sigma_{\rm peak}$ (ALMA position uncertainty) in quadrature. We note that this error is dominated by the JWST position uncertainties because the ALMA uncertainties are small in comparison. }
\end{deluxetable*}

James Webb Space Telescope Near-InfraRed Camera (JWST NIRCam) operates between 0.6 and 5.0 µm. This spectral range is dominated by emission from the SMBH accretion disk for quasars at $z \gtrsim 6$, with the host galaxy contributing only 1\%---5\% of the flux \citep{Fujimoto_2022, Ding_2023, Yue_2024}. Cutouts of the NIRCam observations for each source are shown in Figure \ref{fig:nircam}. All JWST datasets were obtained from the Mikulski Archive for Space Telescopes, detailed in Table \ref{tab:jwst}. The specific observations analyzed can be accessed via \dataset[doi: 10.17909/tyep-h415]{https://doi.org/10.17909/tyep-h415}. In this Table, we list the name of the quasar and the JWST program number for this quasar. For all quasars, we used the standard stage 3 pipeline products delivered from the archive to perform subsequent analysis. JWST observations were obtained using three filters: F115W ($1-1.3$ µm), F200W ($1.7-2.3$ µm), and F356W ($3.1-4.1$ µm). Since these observations were taken at different times, the orientation of the detectors (and consequently the number of visible background stars) varied between filters. For two quasars (J0923$+$0402 and J0244$-$5008), all three filters were used to test for consistency in astrometric accuracy.
As the quasar emission resembles a point source in all filters, and shows no evidence of a significant host galaxy contribution, the quasar coordinates found were in agreement between filters. The filters yielded comparable astrometric accuracy, so the filter with the largest number of field stars was selected for each source. The chosen filter for each source is noted in Table~\ref{tab:jwst}. In fields with abundant background sources, only GAIA sources from the module containing the quasar are included. In sparser fields, sources from both NIRCam modules are used.

\section{Analysis} \label{sec:analysis}

\begin{figure*}
    \centering
    \includegraphics[width=0.96\textwidth]{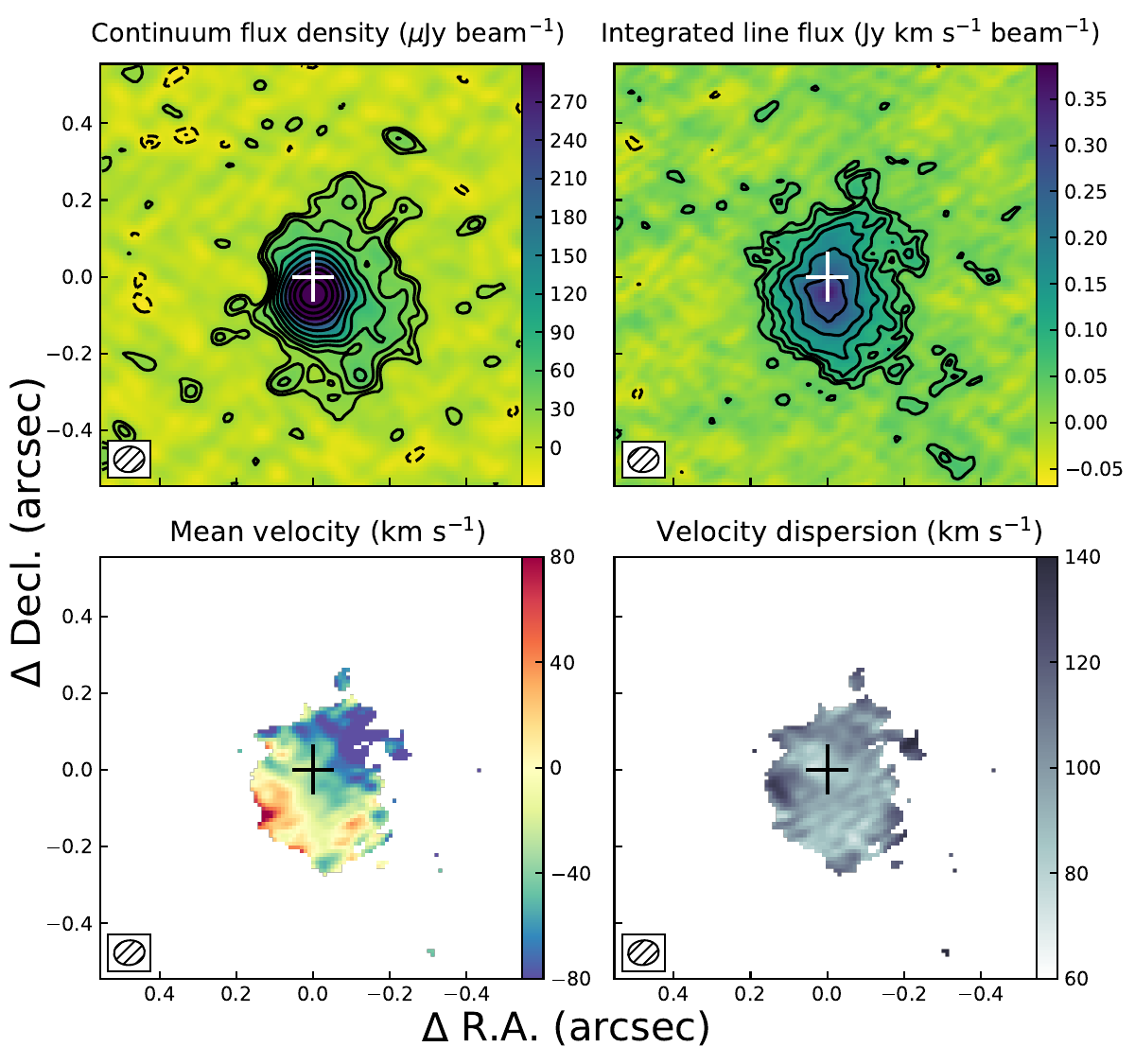}
    \caption{J2054$-$0005 continuum, \CII\ velocity-integrated flux density, mean velocity, and velocity dispersion maps. The cross marks the position of the quasar within the galaxy with error bars as found by comparing JWST and GAIA data (detailed in Section \ref{subsec:astrometry}). The contours in the top two panels start at 2$\sigma$ and increase by powers of $\sqrt{2}$. The synthesized beam is shown in the bottom left corner of each panel.}
    \label{fig:4pan 2054}
\end{figure*}

For each source, we present dust continuum, \CII\ intensity, mean velocity, and velocity dispersion maps in Figure \ref{fig:4pan 2054} (J2054$-$0005 shown in the text, others shown as Figures \ref{fig:4pan 0923}, \ref{fig:4pan 2002}, \ref{fig:4pan 2102}, \& \ref{fig: 4pan 0244} in the Appendix). These plots were produced using the python code \textit{qubefit} \citep{Neeleman_2020}. Integrated flux density maps were made by integrating \CII\ flux across the range of emission (see channel maps, Figure \ref{fig:chanmap 2054}). For the mean velocity and velocity dispersion maps, a Gaussian is fit to the spectrum of each spatial pixel with a signal-to-noise ratio greater than $1 \sigma$. The mean velocity field corresponds to the mean of this Gaussian profile, while the velocity dispersion field is the square root of its variance. The final mean velocity and velocity dispersion plots show only pixels that fall within the $3\sigma$ contour of the \CII\ emission. Consistent with previous studies at $z > 6$ \citep{Venemans_2020, Novak_2020}, we find the \CII\ emission is generally less concentrated and more spatially extended than the continuum.

Continuum and \CII\ flux densities, along with derived luminosities are listed in Tables \ref{tab:cont} and \ref{tab:cii}. Continuum fluxes were measured with circular apertures of diameter $1\farcs5$, while \CII\ fluxes were extracted using circular apertures of diameter $\bm{1\farcs2}$. Several aperture diameters were tested and these were selected to include all the emission of the source, without introducing additional noise. We did not apply a residual scaling correction, because differences between the area of the dirty beam and clean beam were less than 10 \% over all scales of interest. To the formal uncertainties we add, in quadrature, a 10\% uncertainty due to the absolute flux calibration uncertainties. Total Infrared (TIR) luminosities were estimated by modeling the dust emission as an optically thin modified blackbody with a fixed dust temperature of 47 K and an emissivity index of $\beta = 1.6$, based on average properties of high redshift quasars measured in \cite{Beelen_2006}. We apply a correction for Cosmic Microwave Background (CMB) induced suppression of high-redshift flux densities \citep{daCunha_2013}. Uncertainties in TIR luminosities reflect a range of assumed dust temperatures $T_{\rm d}=35-55$ K and emissivity indexes $\beta = 1.2-2.0$. Reported \CII\ line luminosities were calculated following \citet{Solomon_1997}, as implemented in \textit{qubefit} \citep{Neeleman_2020}. 

In general, our measured continuum flux densities and thus dust luminosities are higher, but consistent within the uncertainties, compared to previously reported values \citep{Venemans_2020,Wang_2024}. The higher values for the dust continuum measurement compared to the results reported in \citet{Wang_2024} likely stems from removing the assumption that the continuum emission arises from a Gaussian profile. Our \CII\ luminosities for J0923$+$0402 and J2102$-$1458 are consistent with previous measurements within the reported uncertainties, while the value for J2002$-$3013 agrees to within 20\% of earlier results. Our measurements for J2054$-$0005 yield higher dust and \CII\ luminosities than those reported by \citet{Venemans_2020}, as we resolve an additional emission component not detected in their data. The origin and significance of this excess emission are further discussed in Section~\ref{subsec:galaxy modeling}.

To measure the redshifts for each quasar, we fitted a Gaussian profile to the spatially integrated \CII\ spectrum. The central frequency of the line was converted to redshift assuming a rest-frame frequency of 1900.537 GHz. Uncertainties were taken as the larger of the formal Gaussian fit error or the deviation from the kinematic redshift determined in Section \ref{subsec:galaxy modeling}. Final redshift values are reported in Table \ref{tab:cii}. All redshift values reported here agree with previous measurements, within uncertainty \citep{Farina_2022, Yang_2021, Reed_2019}.

\begin{deluxetable*}{cccccc}
\tabletypesize{\scriptsize}
%\tablenum{1}
\tablecaption{ALMA Quasar Continuum Properties \label{tab:cont}}
\tablehead{
\colhead{Quasar} & \colhead{Peak R.A.}& \colhead{Peak Decl.}& \colhead{$\sigma_{\rm peak}$}& \colhead{$F_{\rm cont}$}& \colhead{$L_{\rm TIR}$}\\
\colhead{   }& \colhead{(h:m:s)}& \colhead{(d:m:s)}& \colhead{(mas)}& \colhead{(mJy)} & \colhead{($10^{12}L_\odot$)}
}
\startdata 
J2054$-$0005& 20:54:06.502& $-$00:05:14.460& 3.3& $3.2 \pm\ 0.7$& $12.5^{+20.7}_{-8.3}$\\
J0923+0402& 09:23:47.122& $+$04:02:54.389& 16.8&  $0.61 \pm\ 0.1$& $2.8^{+4.5}_{-1.8}$\\
J2002$-$3013& 20:02:41.592& $-$30:13:21.580& 4.3&  $2.3 \pm\ 0.7$& $10.7^{+17.4}_{-3.9}$\\
J2102$-$1458& 21:02:19.220 &  $-$14:58:54.030& 6.3& $1.4 \pm\ 0.3$&  $6.3^{+10.3}_{-4.1}$\\ 
J0244$-$5008& 02:44:01.035& $-$50:08:53.740& 6.6&  $0.8 \pm\ 0.2$&  $3.7^{+6.1}_{-2.4}$\\ 
{J0109$-$3047}& 01:09:53.125& $-$30:47:26.320& 4.1& $0.5 \pm 0.1$& $0.4^{+6.3}_{-0.2}$\\
\enddata
\tablecomments{Peak R.A and Peak Decl denote the right ascension and declination of the brightest pixel in the continuum image. $\sigma_{\rm peak}$ is the positional uncertainty on this brightest pixel derived from dividing the beam size by the square root of the signal to noise ratio of this pixel. $F_{\rm cont}$ is the flux density as measured within a circular region of diameter 1.5" centered on the brightest pixel, and $L_{\rm TIR}$ is the total infrared luminosity where the uncertainties re obtained by varying the dust temperature between 35-55K and $\beta$ between 1.2-2. \citep[e.g.,][]{Beelen_2006}.}
\end{deluxetable*}

\begin{deluxetable*}{cccccccccc}
\tabletypesize{\scriptsize}
\tablecaption{ALMA Quasar \CII\ Properties \label{tab:cii}}
\tablehead{
\colhead{Quasar} & \colhead{Peak R.A.}& \colhead{Peak Decl.}& \colhead{$\sigma_{\rm peak}$}&   \colhead{$\rm FWHM_{\text{\CII}}$}& \colhead{$z_{\text{\CII}}$}& \colhead{$F_{\text{\CII}}$}& \colhead{$L_{\text{\CII}}$}\\
\colhead{   }& \colhead{(h:m:s)}& \colhead{(d:m:s)}& \colhead{(mas)} & \colhead{(\kms)}& \colhead{}& \colhead{(Jy~\kms)}& \colhead{($10^{9} L_\odot$)}
}
\startdata 
J2054$-$0005 & 20:54:06.502 & $-$00:05:14.447 & 12.0 & $240\pm11$ & $6.0394\pm0.0004$ & $5.3\pm0.6$ & $5.08\pm0.6$\\  
J0923$+$0402 & 09:23:47.125 & $+$04:02:54.490 & 23.0 & $448\pm64$ & $6.6337\pm0.0005$ & $1.6\pm0.3$ & $1.7\pm0.3$\\
J2002$-$3013 & 20:02:41.591 & $-$30:13:21.586 & 20.0 & $240\pm33$ & $6.6873\pm0.0002$ & $1.60\pm0.24$ & $1.8\pm0.3$\\ 
J2102$-$1458 & 21:02:19.221 & $-$14:58:54.030 & 20.9 & $270\pm44$ & $6.6655\pm0.0007$ & $1.24\pm0.21$ & $1.37\pm0.23$\\ 
J0244$-$5008 & 02:44:01.035 & $-$50:08:53.740 & 16.4 & $240\pm46$ & $6.7307\pm0.0004$ & $1.3\pm0.3$ & $1.5\pm0.3$\\ 
J0109$-$3047& 01:09:53.125& $-$30:47:26.324& 26.0& $360 \pm 39$& $6.7903 \pm 0.0005$& $1.3 \pm 0.2$& $1.4 \pm 0.2$\\
\enddata
\tablecomments{
Peak R.A and Peak Decl denote the right ascension and declination of the brightest pixel in the \CII\ moment 0 image. $\sigma_{\rm peak}$ is the positional uncertainty on this brightest pixel derived from dividing the beam size by the square root of the signal to noise ratio of this pixel. $\rm FWHM_{\text{\CII}}$ is the Full Width Half Max of the \CII\ line and $z_{\text{\CII}}$ is redshift, based on fitting a Gaussian to the spectrum from a 0\farcs6 aperture. Redshift uncertainties were found based on the difference from redshift computed based on the central velocity found in kinematic modeling (see Section \ref{subsec:galaxy modeling}) or the uncertainty of the gaussian fit, whichever was larger. $F_{\text{\CII}}$ is the \CII\ Flux Density from integrating the Gaussian fit, and $L_{\text{\CII}}$ is the corresponding luminosity, following \cite{Carilli_Walter_2013}.} 
\end{deluxetable*}

\begin{deluxetable*}{ccccccccc}
\tabletypesize{\scriptsize}
%\tablewidth{0pt} 
%\tablenum{1}
\tablecaption{Kinematic Modeling Parameters and Fit Statistics\label{tab:fit info}}
\tablehead{
\colhead{Quasar} & \colhead{$i$} & \colhead{$\sigma_v$} & \colhead{$\chi^2$} & \colhead{$v_{\rm rot}$} & \colhead{$R_{\rm d}$} & \colhead{$M_{\rm dyn}$} & \colhead{$M_{\rm BH}$}\\
\colhead{   } & \colhead{($deg$)} & \colhead{(\kms)} & \colhead{  } & \colhead{(\kms)} & \colhead{(kpc)} & \colhead{($10^{10}M_\odot$)}& \colhead{($10^9 M_{\odot}$)}
}
\startdata 
J2054$-$0005 & 11.6 & $159.04_{-2.12}^{+2.20}$ & 1.5 & $526.1_{-27.8}^{+28.3}$ & $1.301_{-0.017}^{+0.016}$ & $8.9_{-0.9}^{+0.9}$ & $2.2\pm0.3$\\  
{J0923$+$0402}& $-$& $-$& $-$& $-$& $-$& $-$& $-$\\
J2002$-$3013 & 27.5 & $122.23_{-4.19}^{+4.3}$ & 1.13 & $314.6_{-10.8}^{+9.9}$ & $0.86_{-0.03}^{+0.03}$ & $1.99_{-0.15}^{+0.15}$ & $1.6\pm0.3$\\ 
J2102$-$1458 & 13.7 & $143.87_{-4.8}^{+5.16}$ & 0.94 & $222.8_{-35.5}^{+36.3}$ & $0.77_{-0.03}^{+0.03}$ & $0.9_{-0.3}^{+0.3}$ & $0.7\pm0.11$\\ 
J0244$-$5008 & 14.16 & $156.36_{-5.9}^{+7.12}$ & 1.24 & $604.8_{-132.5}^{+163.4}$ & $0.67_{-0.03}^{+0.03}$ & $5.74_{-2.5}^{+3.11}$ & $1.2\pm0.4$ \\
{J0109$-$3047}& $-$& $-$& $-$& $-$& $-$& $-$& $-$& \\
\enddata
\tablecomments{
Parameters used for, and obtained from, the kinematic modeling process using the thin disk model in \textit{qubefit}. $i$ is the inclination angle, measured from the \CII\ moment-zero image and fixed in the fitting process. $\chi^2$ is the reduced chi squared of the fit. $\sigma_v$ is the velocity dispersion, $v_{\rm rot}$ is the rotational velocity, and $R_{\rm d}$ is the galaxy radius, all of which are outputs of the fit. The dynamical mass was calculated from the given information, following \citet{Neeleman_2021} and described in Section \ref{subsec:galaxy modeling}. $M_{\rm BH}$ is the black hole mass as measured from the \ion{Mg}{2}\ line (J2054$-$0005 from \citet{Farina_2022}; J2002$-$3013 and J2102$-$1458 from \citet{Yang_2021}; and J0244$-$5008 from \citet{Reed_2019}). J0109$-$3047 values are not included because the kinematic modeling performed in \cite{Meyer_2023} determined the source to be best fit by a dispersion-dominated model, which has different parameters.}
\end{deluxetable*}

\subsection{Astrometry} \label{subsec:astrometry}

As mentioned in Section \ref{sec:jwst}, at the wavelength coverage of JWST/NIRCam, the accretion disk emission of a $z\gtrsim 6$ quasar dominates the spectrum for these sources \citep{Fujimoto_2022, Ding_2023, Yue_2024}. As such, we can use these data to find accurate coordinates for the position of the SMBH. Because we use NIRCam stage 3 images, all images have their astrometric solution tied to the GAIA reference frame. In order to provide an observational uncertainty on this astrometric solution, we identified a set of isolated and unsaturated background stars which have been cataloged by GAIA's third data release. The median number of background stars for our sample of quasars was 20, including stars in both NIRCam modules.   

The offsets in Right Ascension and Declination between the JWST positions of background stars and their corresponding GAIA coordinates were tracked for each source. We measure the position of background stars both by brightest pixel and by a 2D Gaussian fitting routine. The coordinates and corresponding offsets from GAIA show no significant difference depending on method. While quasar positions are measured both by brightest pixel and by a flux-weighted-average method, the values listed in Table \ref{tab:jwst} are those using the brightest pixel method. We take as the quasar's positional uncertainty the standard deviation of the measured offsets for all background stars. The standard deviation tended to decrease as more background stars were considered, and plateau once around 25 background stars had been included. This meant for J2002$-$3013 and J2054$-$0005, all stars in the NIRCam module with the quasar were included, but not all stars on the second module were. For all other sources we include all available background sources. 

 \begin{figure}
    \centering
    \includegraphics[width=1\linewidth]{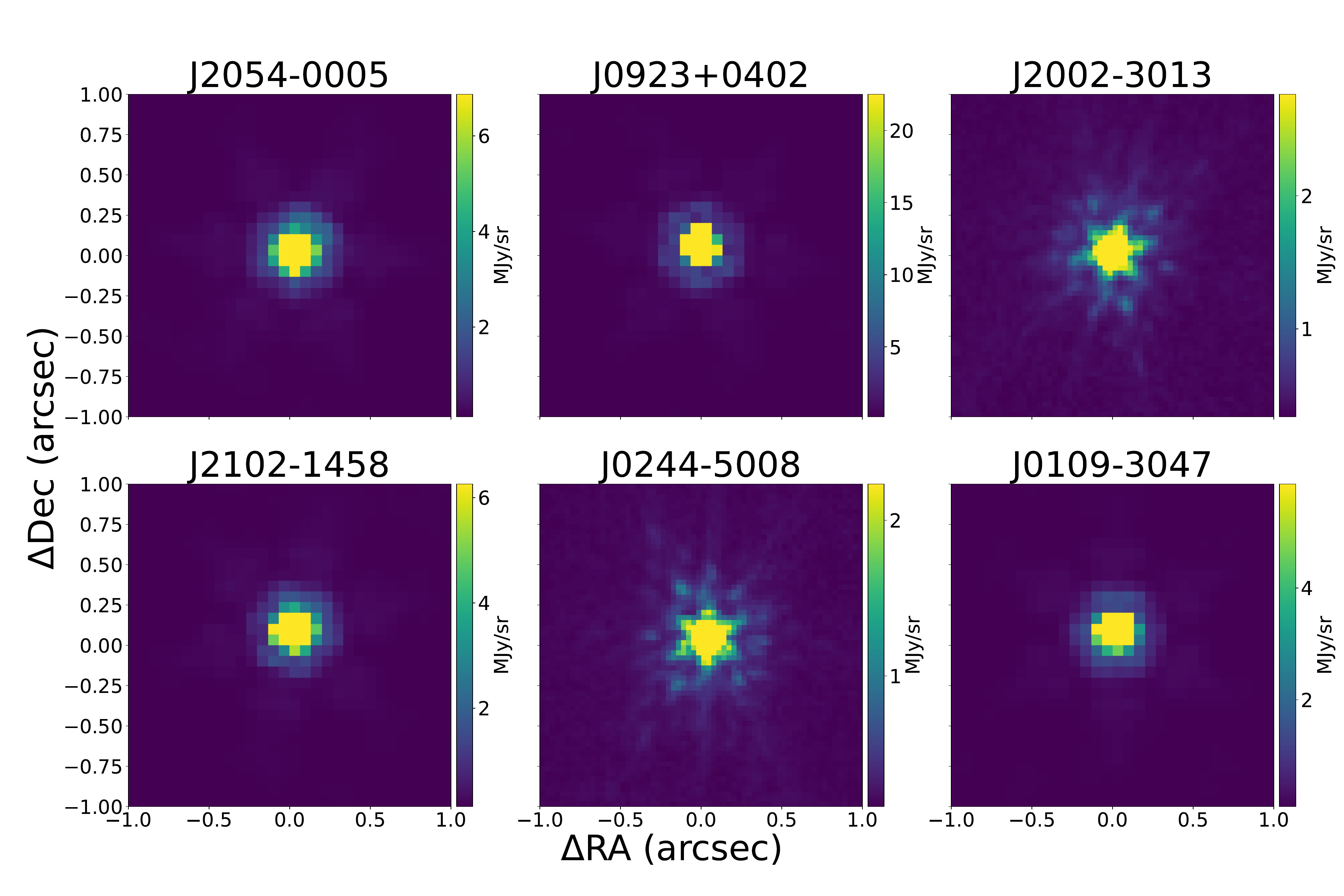}
    \caption{JWST NIRCam images of each source in our sample, shown in the filter used for GAIA comparison and coordinate measurements (see Table \ref{tab:jwst}). The axes are centered on the measured position of each source.} 
    \label{fig:nircam}
\end{figure}

The resulting quasar positions, along with their associated uncertainties, are listed in Table \ref{tab:jwst}. Figure \ref{fig:nircam} shows the JWST NIRCam images of each source, where the axes are centered on their measured positions. These positions are also reflected in the location and size of each cross in Figure \ref{fig:4pan 2054}. Compared to the previously published coordinates based on ground-based optical imaging data \citep{Venemans_2020}, the JWST positions yield improved initial astrometric accuracy due to its higher spatial resolution.

\begin{figure*}
    \centering
    \includegraphics[width=0.74\linewidth]{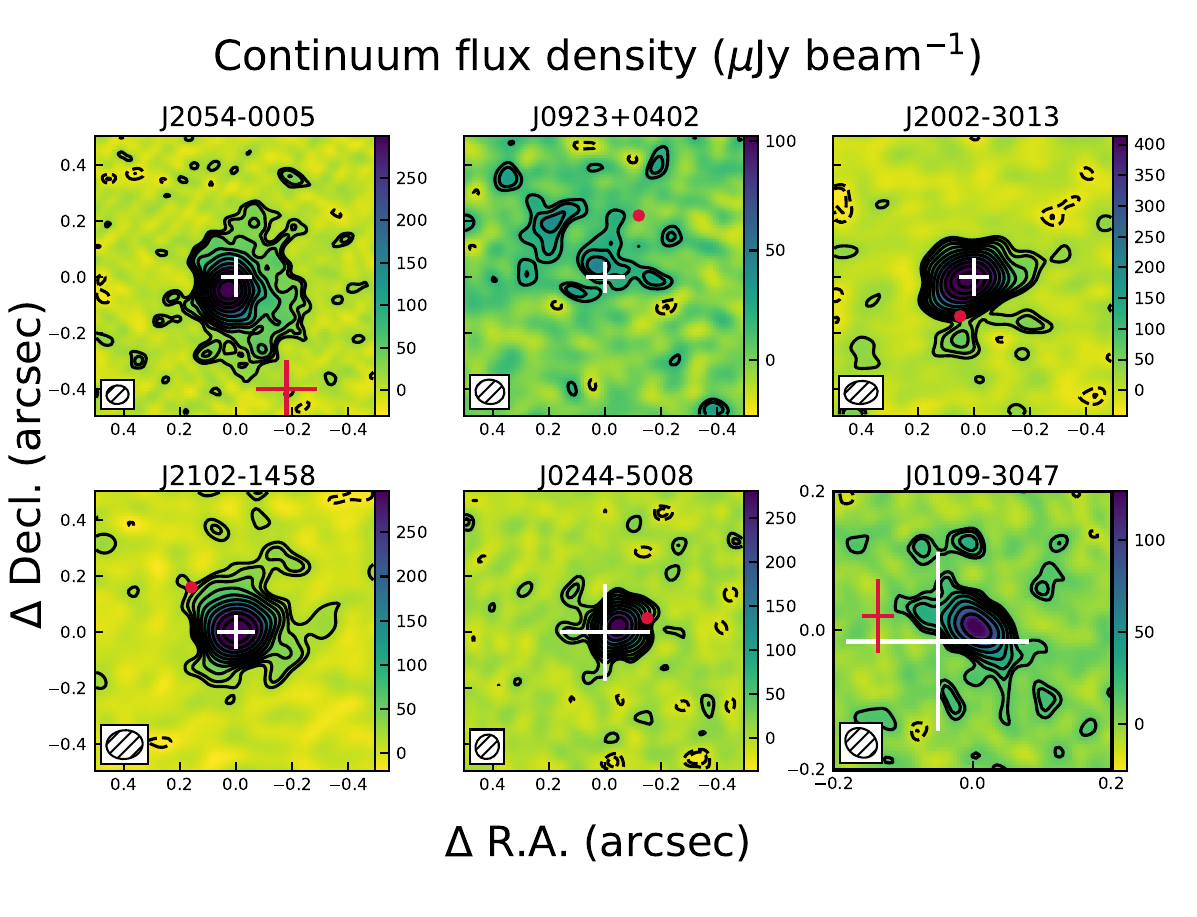}
   \caption{Continuum imaging of all six sources in the sample. Previous optical position of the quasars are noted in red. JWST positions (see Section \ref{sec:jwst}) of the quasars are shown in white. For sources lacking reported uncertainties in the literature for their previous coordinates, the positions are marked with a dot. We note that the spatial scaling for J0109$-$3047 is different, for the clearest possible view of the galaxy. Contours and annotations are similar to Figure \ref{fig:4pan 2054}.}
    \label{fig:cont 6pan}
\end{figure*}

\begin{figure*}
   \centering
    \includegraphics[width=0.74\linewidth]{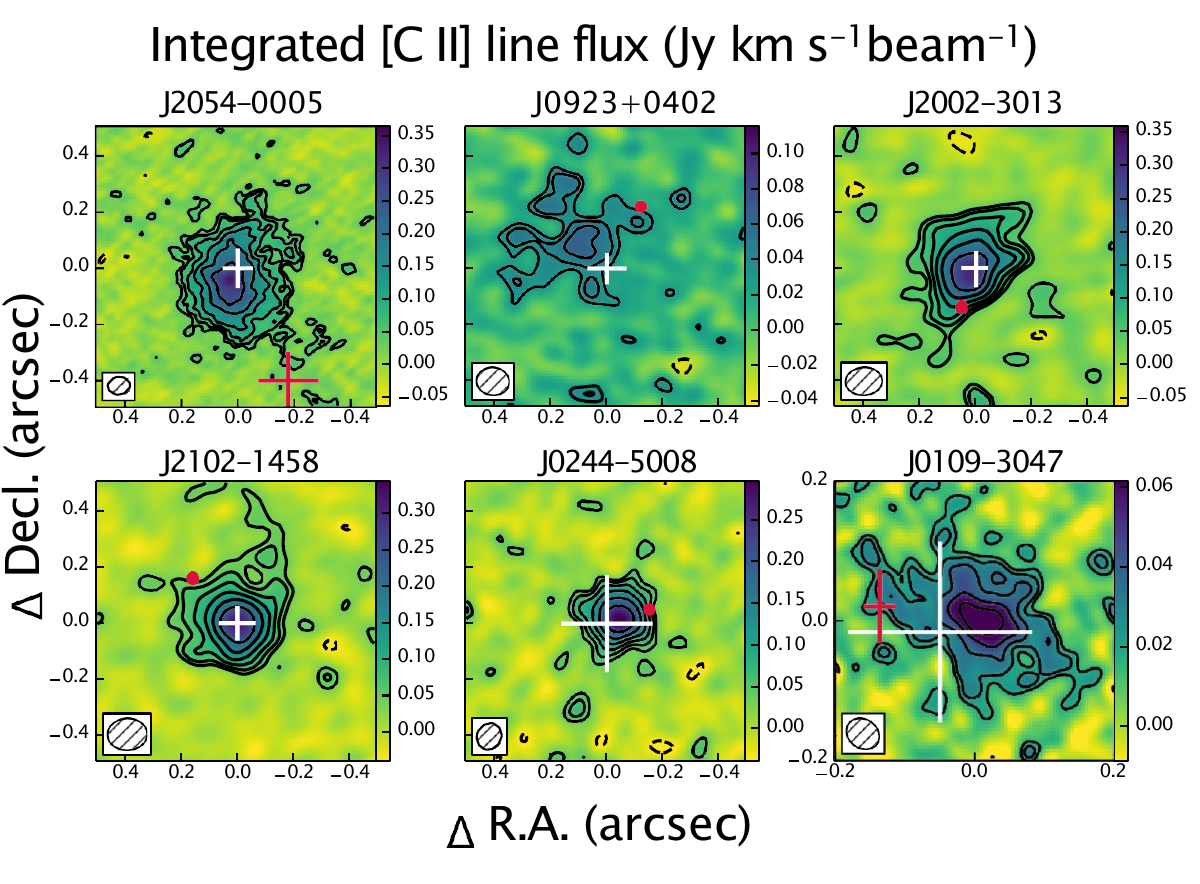}
    \caption{\CII\ imaging of all six sources in the sample. Contours and annotations are similar to Figure \ref{fig:cont 6pan}.}
    \label{fig:mom 6pan}
\end{figure*}

Figures \ref{fig:cont 6pan} and \ref{fig:mom 6pan} present the continuum and velocity-integrated \CII\ maps with the JWST GAIA-corrected position in white and the previous position based on optical ground-based observations in red. Table \ref{tab:jwst} details the spatial offset between the new JWST coordinates and the centers of continuum and \CII\ emission, respectively, where the emission centers are defined by the brightest pixel. The coordinates of the positional centroid for each host galaxy's continuum and \CII\ were also computed (using a flux-weighted average), and found to fall within the positional error bar on the brightest pixel coordinates (see Tables \ref{tab:alma obs} and \ref{tab:cii}) in most cases. We chose to use the brightest pixel coordinates because it better accounts for any asymmetries or structural irregularities in the galaxy morphology. The uncertainty on this position was computed by dividing the beam size by the square root of the signal to noise ratio of the brightest pixel. The offsets between SMBH and host galaxy are within 1$\sigma$ uncertainties in all but one (J0923$+$0402) case. Thus, most sources are located within the central $\sim 400$ pc of their host galaxy, with some observations constraining offsets to within 100 pc of the galaxy center.  We note that the source with the most significant offset is J0923+0402. However, this source is only weakly detected in \CII, and will require more integration time to yield a reliable measurement of the center of the \CII\ and dust continuum emission. We therefore do not interpret this offset as significant. For J0109$-$3047, the source discussed in \citet{Meyer_2023}, the offset between the black hole and galaxy center was within 1 $\sigma$\ uncertainty. However, the uncertainties on this measurement are large, because there are very few GAIA sources in the background of the JWST imaging. Imaging with a larger field-of-view would help reduce the uncertainties on this source. The previous position measurement for this source was made using the Visible and Infrared Survey Telescope for Astronomy (VISTA), which is an order of magnitude worse in resolution than JWST NIRCam \citep{Meyer_2023}. 

Recent JWST studies of $z \gtrsim 6$ quasars have reported apparent spatial offsets between the SMBH and the host galaxy emission, which may be attributed to dust obscuration in the rest-frame optical and near-infrared \citep{Ding_2023, Yue_2024}. Surveys of \CII~ emission in quasar host galaxies at $z \gtrsim 4$ have shown that [C II] tends to trace regions that are heavily obscured in the rest-frame optical \citep{Fudamoto_2021, Pozzi_2024}. Our \CII-based measurements show no significant offsets, suggesting that the \CII~ emission is largely unaffected by dust. This comparison implies that the offsets observed in JWST imaging are likely driven by dust obscuration, and that SMBHs reside at the centers of their host galaxies.

\subsection{Kinematic Modeling of the Host Galaxy}\label{subsec:galaxy modeling}

\begin{figure*}
    \centering
    \includegraphics[width=\textwidth]{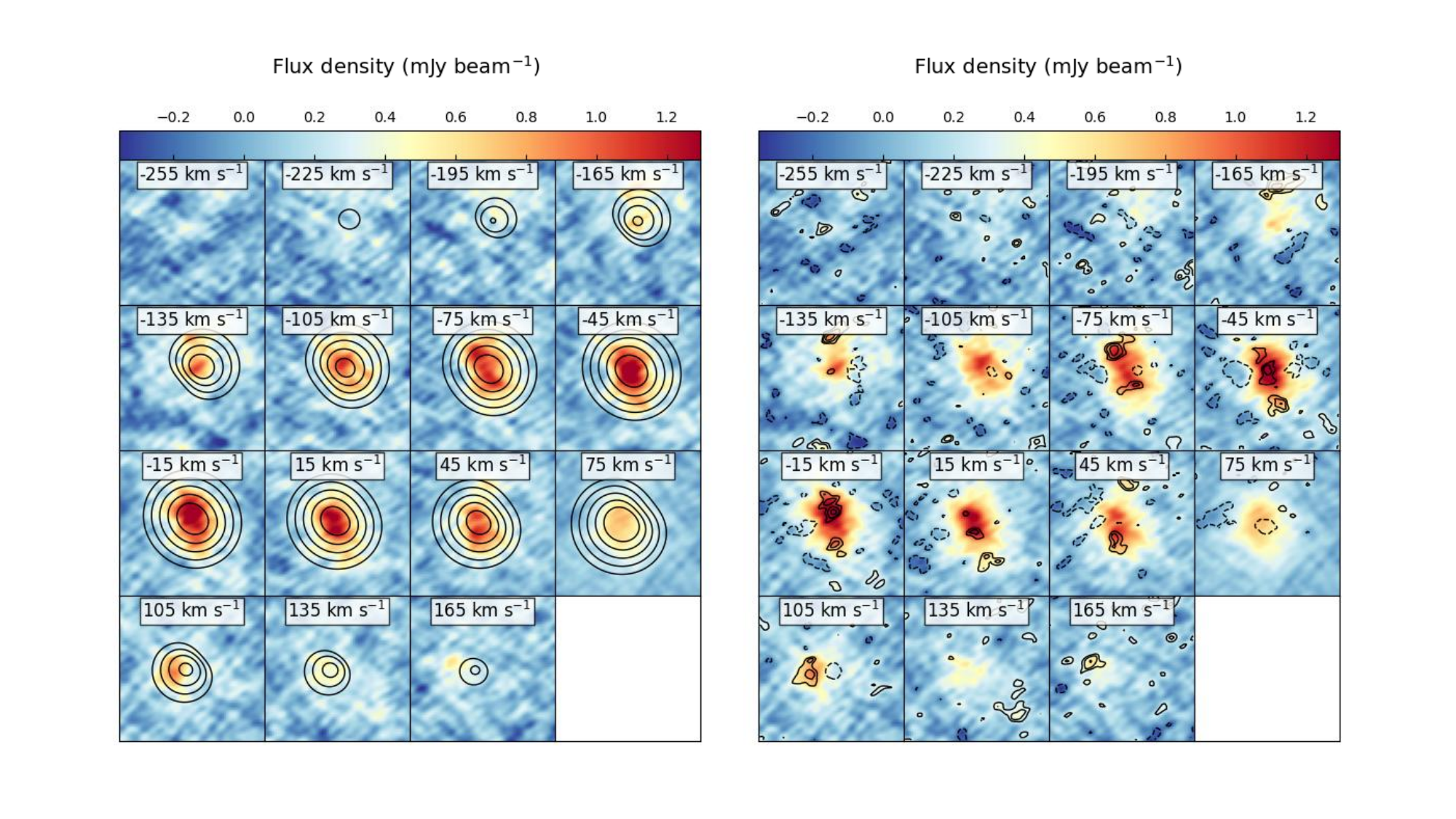}
    \caption{J2054$-$0005 Channel Maps from the \textit{qubefit} thin disk model. The left panel shows \CII\ data at each 30 \kms\ increment with contours of the model. The right panel shows \CII\ emission at each 30 \kms\ increment with residual contours. Both sets of contours start at $2 \sigma$ and increase by factors of $\sqrt{2}$.}
    \label{fig:chanmap 2054}
\end{figure*}

The host galaxies of four of the quasars in our sample were modeled using the \textit{qubefit} code \citep{Neeleman_2020}, which employs a Bayesian Markov Chain Monte Carlo (MCMC) approach to estimate the posterior distributions of model parameters. We did not fit quasar J0923+0402 due to insufficient signal-to-noise ratio in the individual channels. J0109$-$3047 was also excluded, as this analysis has already been presented in \citet{Meyer_2023}. For the fitting, we adopted a user-defined thin disk model with an exponential intensity profile and a constant rotational velocity as described in detail in \citep{Neeleman_2021}. The results are shown in Figure \ref{fig:chanmap 2054} (J2054$-$0005 shown in the text, others available as Figures \ref{fig:chanmap 2002}, \ref{fig: chanmap 2102}, \& \ref{fig:chanmap 0244} in the Appendix). Due to the nearly face-on orientation of most sources, the inclination angle is difficult to constrain using \texttt{qubefit} alone. For J2054$-$0005, J2002$-$3013, and J2102$-$1458, we estimate their inclination, by fitting an ellipse convolved with the synthesized beam to the continuum image and then estimate an inclination under the assumption the galaxy is a circular disk. We chose to fix inclinations in the \textit{qubefit} fitting process, because without these constraints the rotational velocities do not have an upper bound. Table \ref{tab:fit info} lists the adopted inclinations, $\chi^2$ values of each fit and velocity dispersion values, which are consistent with \cite{Neeleman_2021}. 

\begin{figure}
    \centering
    \includegraphics[width=\linewidth]{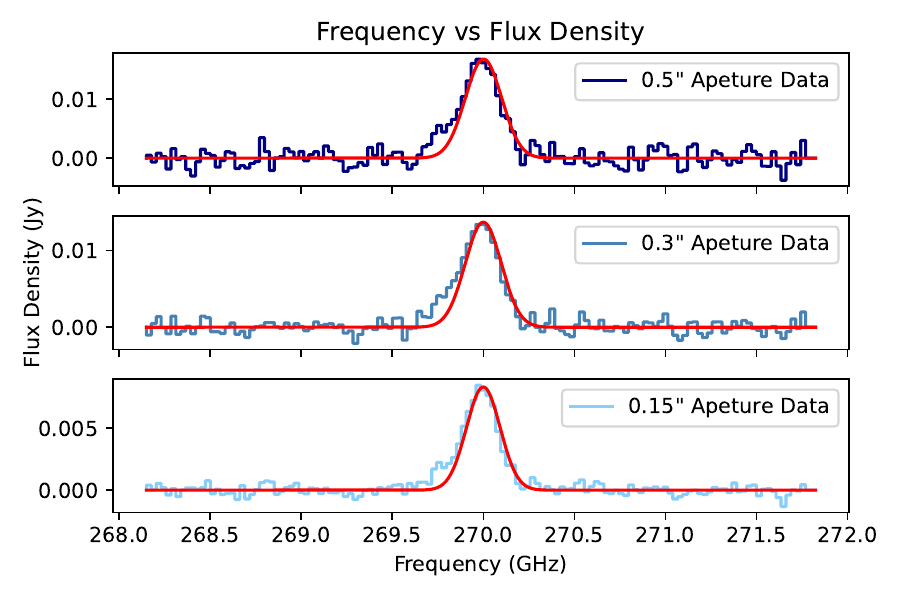}
    \caption{\CII\ spectrum of J2054$-$0005 extracted from three different circular apertures (radii of $0\farcs5$, $0\farcs3$, and $0\farcs15$). All apertures are centered on the brightest pixel of the velocity-integrated \CII\ emission map. We note a low frequency deviation from a Gaussian profile in all three apertures at $\approx$269.75 GHz (+250 \kms). The red models are Gaussian profiles which are fit to the right half ($\geq 269.9$ GHz) of the spectrum, to clarify the deviation at 269.75 GHz.}
    \label{fig:2054 nongauss}
\end{figure}

For J2054$-$0005, we identify a notable deviation from a simple rotating disk model. A residual feature in the north part of the galaxy shows excess emission compared to the model at 3.6$\sigma$ in the 15 \kms\ channel and 3.7$\sigma$ in the $-45$ \kms\ channel. The bottom right residual is less significant, with deviations at the 2.5$\sigma$ and 2.4$\sigma$ level for the 45 \kms\ and 75 \kms\ channels, respectively. In addition, spectra of this source, shown in Figure \ref{fig:2054 nongauss}, exhibit a non-Gaussian shape, with an excess near 269.75 GHz or $\approx$250 \kms\ from the nominal redshift of this galaxy. This feature is detected consistently across three concentric circular apertures centered on the center of J2054$-$0005, suggesting that it is not confined to a single off-center spatial region. This makes it unlikely that the feature arises from a companion galaxy. An alternative explanation is a \CII\ outflow from the central source. Such an outflow could naturally account for the spectral excess, and its uniform appearance across the apertures.

Residual features are also apparent in J2102$-$1458, particularly in the top left, with excess emission of 4.6$\sigma$ and 3.8$\sigma$ in the 15 \kms\ channel and 105 \kms\ channel, respectively. These regions of excess emission could arise from star-forming regions within the disk of the galaxy or denser regions of gas that is able to emit more efficiently in \CII. For the remaining two galaxies, the thin disk model provides a sufficient fit to the data, and no strong excess emission is found, showing that these galaxies emit relatively smoothly on these spatial scales.

For all sources we determined the kinematic redshift from the central velocity, which is a free parameter in the thin disk model that we adopted. These values were used to define the redshift uncertainties in Table \ref{tab:cii} as detailed in Section \ref{sec:analysis}.

From the kinematic fitting we are able to derive dynamical masses using the approach described in \citet{Neeleman_2021}. To be specific, we derive a dynamical mass assuming a spherically symmetric mass distribution, in which case the dynamical mass, $M_{\rm dyn}$ is related to the rotational velocity, $v_{\rm circ}$ and radius, $R$ via:

\begin{equation}
M_{\rm dyn} = \frac{v_{\rm circ}^2}{G} R = 2.33 \times 10^5 v_{\rm circ} R.
\end{equation}

where in the last equality $v_{\rm circ}$ is in \kms, R is the radius in kpc, and the returned $M_{\rm dyn}$ is in $M_{\odot}$. As we do not have the spatial resolution to distinguish circular and random motions, we assume all motion is circular and use $v_{\rm rot}$ from the kinematic model as our circular velocity. The resulting dynamical masses (Table \ref{tab:fit info}) are on the order of $\sim$$10^{10} M_{\odot}$, which is consistent with earlier estimates of high-z galaxies \citep{Neeleman_2021, Wang_2024}. \cite{Neeleman_2021} placed a lower bound of $2.9 \times 10^{10} M_{\odot}$ on the dynamical mass of J2054$-$0005, which is in agreement with our measurement. Similarly, our dynamical mass estimate for J2002$-$3013 result aligns with the dispersion-corrected value of $2.5^{+0.31}_{-0.10} \times 10^{10} M_{\odot}$ reported by \cite{Wang_2024}, though their uncorrected estimate is higher at $5.7^{+0.69}_{-0.22} \times 10^{10} M_{\odot}$.

\section{Discussion} \label{sec:discussion}

\subsection{Offsets} \label{subsec: offsets}

Simulations predict that mergers of black holes with misaligned spins can cause the resultant SMBH to exhibit an observable position or velocity offset \citep{Campanelli_2007, Blecha_2016}. At high redshift ($z \gtrsim\ 6$), however, such recoil events are expected to be rare. The most massive SMBHs likely lack equal-mass companions, and gas-rich mergers are predicted to align black hole spins and increase central escape velocities, both of which reduce the likelihood and magnitude of recoils. Nevertheless, \cite{Meyer_2023} reported that QSO J0109$-$3047 was not found at the center of the \CII\ or continuum emission of its host galaxy. Figures \ref{fig:cont 6pan} and \ref{fig:mom 6pan} show ALMA continuum and \CII\ imaging, respectively, of our sample of six quasars. Each source has its previously published position marked with a red cross/dot and our JWST GAIA-corrected position marked with a white cross. All six of these quasars are found to be centrally located within their host galaxies, within a 1$\sigma$ uncertainty, compared to both their continuum and \CII\ emission centers (a result recently confirmed in infrared wavelengths in \cite{Ding_2025}). This holds true also for the GAIA-corrected JWST position of J0109$-$3047, although we acknowledge with large uncertainties due to the lack of sufficient stars to align the JWST image. However, the nominal JWST position of the quasar is in much better agreement with the center of the galaxy as determined from either \CII\ or the dust emission with respect to the optical position as found in \cite{Meyer_2023}. We also note a significant disagreement of the JWST-GAIA corrected and optical position for J2054$-$0005, which is particularly interesting because the optical coordinates are GAIA-corrected. The previous position comes from ground-based surveys which may have different visible background stars than corresponding JWST imaging or demonstrate a systematic offset due to differential chromatic refraction effects \citep{Kaczmarczik_2009, Venemans_2020}. We also note that the optical (rest-frame UV) emission may be much fainter due to dust attenuation, making the centroid measurement more uncertain and potentially contributing to the found discrepancy between JWST and optical centroids \citep{Nedkova_2024}.

The lack of measurable positional offsets in our sample supports theoretical expectations that recoil events are either intrinsically rare or short-lived in gas-rich, high-redshift galaxies \citep{Blecha_2016, Dunn_2020, Dong-Paez_2025}. We note the caveat that an offset AGN's surrounding gas may disperse, causing it to no longer appear as a quasar and thus not be detected in our sample \citep{Prochaska_2014, Hopkins_2008}. It is also possible that these sources exhibit a velocity offset, which we do not attempt to quantify due to large systematic velocity shifts ($\sim 500$ \kms) in redshift measurements between the \ion{Mg}{2} and \CII\ lines \citep{Meyer_2019}. Nevertheless, our sample does not exhibit a significant positional offset between host galaxy and quasar. Our results are consistent with models in which cold gas inflows during wet mergers induce black hole spin alignment  \citep{Campanelli_2007, Volonteri_2010, Dunn_2020}. This gas may further suppress recoil events by increasing the escape velocity of the central black holes, and creating sufficient dynamical friction to counter any black hole from escaping \citep{Volonteri_2010, Blecha_2016, Dunn_2020}. The observed alignment of the quasar with the center of the host galaxy suggests that SMBHs in massive, $z \geq 6$ galaxies are unlikely to experience large recoils due to mergers or this offset due to a recoil event is short-lived. On a similar note, despite the fact that we know that these systems are formed from continued accretion and mergers from the cosmic web, their dense gas, traced either via dust of \CII\ emission, remains remarkably undisturbed and continue to surround the SMBH. This either suggests that mergers do not disturb this inner region, or the dynamical timescale for gas settling is sufficiently short. Numerical works have suggested that this disk-like structure may also be explained by the inefficiency of the supernova in expelling gas in the most massive galaxies \cite{Lupi_2019}. 

This study, while sensitive to large-scale quasar offsets, does not have the resolution to distinguish offsets that occur within $\sim 400$ pc of the galaxy center for all sources. Detecting small-scale displacements would require improved astrometric precision in the near-infrared, as well as higher-resolution ALMA imaging to constrain the sphere of influence of the black hole.

\subsection{Kinematic Modeling} \label{subsec:discuss kinematics}

Larger surveys that have modeled the kinematics of high redshift quasar host galaxies have observed a wide range of morphologies, including active mergers, disk-like galaxies, and dispersion-dominated systems \citep{Neeleman_2021}. In our sample, four quasars--- J2054$-$0005, J2002$-$3013, J2102$-$1458, and J0244$-$5008--- are well modeled by a thin disk (model detailed in Section \ref{subsec:galaxy modeling}. This is consistent with previous lower-resolution modeling of J2054$-$0005, which also supported a thin disk model \citep{Neeleman_2021}. The agreement with thin disk models suggests these galaxies are in a post-merger or secular phase of evolution, rather than an active merger.

\begin{figure}
    \centering
    \includegraphics[width=1\linewidth]{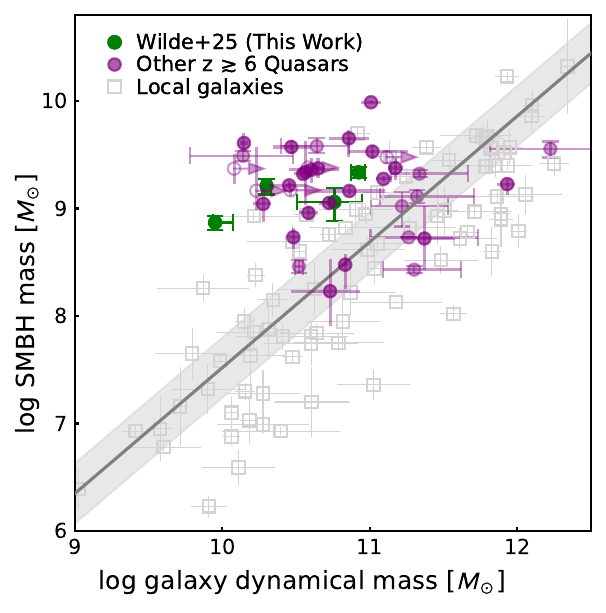}
    \caption{Black hole mass versus dynamical mass for our sample of $z \gtrsim\ 6$ quasars. Other high redshift quasars plotted are sourced from \cite{Neeleman_2021, Wang_2024, Yue_2024}. Also shown is the same relation for local galaxies \citep{Kormendy_Ho_2013, DeNicola_2019}, and the gray line fit derived by \cite{Kormendy_Ho_2013} where the shaded region is the 1$\sigma$ scatter about this correlation.}
    \label{fig:mass ratios}
\end{figure}

In Section \ref{subsec:galaxy modeling}, we present dynamical mass estimates for these four sources, including the first such measurements for J2102$-$1458 and J0244$-$5008. Black hole masses have been estimated for all sources based on the \ion{Mg}{2} line \citep{Reed_2019, Yang_2021, Farina_2022}. In Figure \ref{fig:mass ratios}, we plot the log ratio of SMBH mass to dynamical mass alongside the broader samples in \cite{Neeleman_2021, Wang_2024, Yue_2024} as well as the local AGN sample described in \citet{DeNicola_2019}. Our results confirm that these quasars, like other high-redshift sources, lie above the local $M_{\rm BH} - M_{\rm bulge}$ relation. This suggests that, in agreement with past studies \citep{Walter_2003, Ho_2007, Willott_2015, Venemans_2016, Shao_2017, Izumi_2018, Decarli_2018, Pensabene_2020}, high redshift SMBHs traced by luminous quasars are hosted, on average, by smaller galaxies compared to their local counterparts. We recognize, however, that these sources were discovered because of their bright UV continuum, which likely biases our selection towards sources with exceedingly massive black holes \citep{Lauer_2007}.  

These higher resolution ALMA imaging are important for reliable kinematic modeling, particularly for constraining the disk inclination. Since inclination is the dominant source of uncertainty, this significantly improves the fidelity of the dynamical mass estimates. In addition, the higher resolution ALMA imaging allows for a more reliable identification of merger signatures, as well as a way to select potential sources that exhibit a SMBH recoil. If positional SMBH offsets are observable in some cases, analyzing the differences in host galaxy kinematics could provide further insight into the dynamical conditions that give rise to recoil events. The lack of observable offsets in these systems is consistent with their absence of merger signatures, if we assume that the time-scales for observing evidence of a merger and of a recoil event are similar. However, larger samples of galaxies are needed to test this assumption.

\section{Summary \& Conclusions} \label{sec:conc}
We have presented ALMA dust continuum and \CII\ emission imaging of five $z \gtrsim 6$ quasar host galaxies: J2054$-$0005, J0923$+$0402, J2002$-$3013, J2102$-$1458, \& J0244$-$5008. Updated redshifts based on the \CII\ emission line have been derived for all sources in this paper. Previous observations of J2054$-$0005 show evidence of a disk galaxy structure, which was confirmed at this higher resolution despite evidence for some excess emission. The other sources also appear to be disk galaxies, as demonstrated by kinematic modeling. 

We further determine that all six sources (including J0109$-$3047, which was previously imaged in \cite{Meyer_2023}) do not exhibit a spatial offset between the optical/near-infrared position of the quasar and the position of the host galaxy as determined from \CII\ and dust continuum emission. This lack of offset in our sample, compared to displacements which have been observed in rest-frame optical JWST imaging, suggests that those offsets are likely caused by dust obscuration and do not reflect a genuine physical separation between the SMBH and its host galaxy \citep{Ding_2025}. Using new GAIA-corrected JWST observations, we find no significant offset in J0109$-$3047, although the uncertainties on this measurement are large due to the lack of sufficient background stars within the JWST image. Future near-infrared adaptive optics-corrected large field of view imaging would serve to reduce uncertainties on this measurement. These observations showcase the synergy between JWST imaging and high resolution ALMA observations in order to constrain the correlation between black holes and galaxies at $z \gtrsim 6$.

\begin{acknowledgements}
This paper makes use of the following ALMA data:\\
ADS/JAO.ALMA\#2018.1.00908.S (PI: Walter, F.),\\ ADS/JAO.ALMA\#2018.1.01188.S (PI: Wang, F.),\\ ADS/JAO.ALMA\#2019.1.00672.S (PI: Fujimoto, S.),\\ ADS/JAO.ALMA\#2019.1.01025.S (PI: Wang, F.),\\ ADS/JAO.ALMA\#2021.1.00934.S (PI: Yang, J.).\\
ALMA is a partnership of ESO (representing its member states), NSF (USA) and NINS (Japan), together with NRC (Canada), MOST and ASIAA (Taiwan), and KASI (Republic of Korea), in cooperation with the Republic of Chile. The Joint ALMA Observatory is operated by ESO, AUI/NRAO and NAOJ. The National Radio Astronomy Observatory is a facility of the National Science Foundation operated under cooperative agreement by Associated Universities, Inc.

This work is based [in part] on observations made with the NASA/ESA/CSA James Webb Space Telescope.The data were obtained from the Mikulski Archive for Space Telescopes at the Space Telescope Science Institute, which is operated by the Association of Universities for Research in Astronomy, Inc., under NASA contract NAS 5-03127 for JWST. These observations are associated with program \#2078 and \#1813.

RAM acknowledges support from the Swiss National Science Foundation (SNSF) through project grant 200020\_207349.
\end{acknowledgements}

\facilities{ALMA, JWST/NIRCam}

\software{Astropy \citep{Astropy_2013, Astropy_2018, Astropy_2022}, CASA \citep{casa}, Matplotlib \citep{matplotlib}, Qubefit \citep{Neeleman_2020}.}

\bibliography{bh-offsets}{}
\bibliographystyle{aasjournal}
\newpage
\section{Appendix} \label{sec:tab&fig}

Continuum and moment 0-2 maps (analogous to Figure \ref{fig:4pan 2054}) for J0923+0402, J2002$-$3013, J2102$-$1458, and J0244$-$5008 are included here. Channel maps (analogous to Figure \ref{fig:chanmap 2054}) for J2002$-$3013, J2102$-$1458, and J0244$-$5008 are also shown.

\begin{figure}[h]
    \centering
    \includegraphics[width=0.96\textwidth]{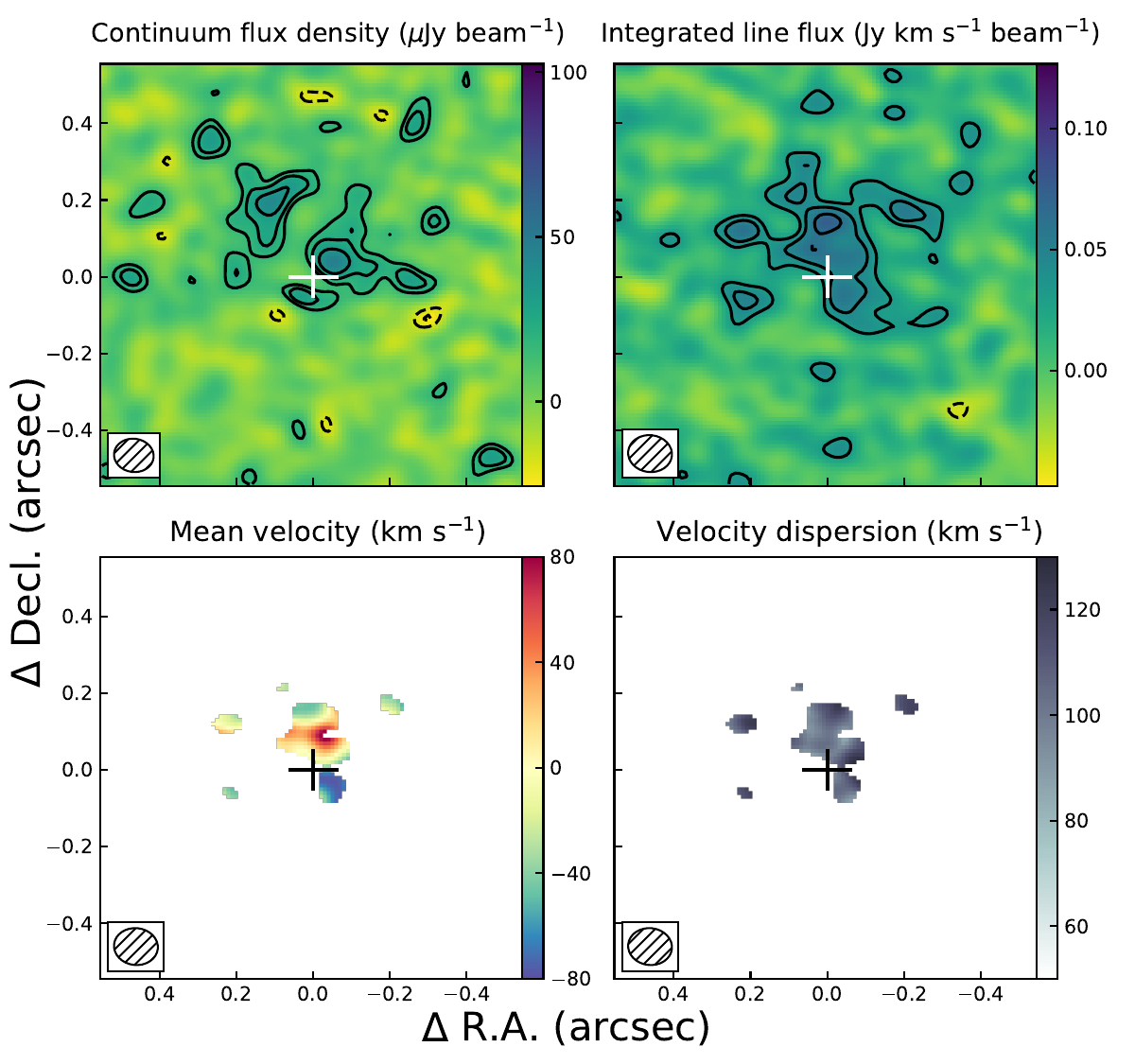}
    \caption{J0923$+$0402 continuum, \CII\ Integrated Intensity, Mean Velocity, and Velocity Dispersion maps. The cross marks the position of the quasar within the galaxy with error bars, as found by comparing JWST and GAIA data (detailed in Section \ref{subsec:astrometry}). The contours in the top two panels start at 2 $\sigma$ and increase by factors of $\sqrt{2}$. The synthesized beam is shown in the bottom left corner of each panel.}
    \label{fig:4pan 0923}
\end{figure}
\begin{figure*}
    \centering
    \includegraphics[width=0.96\textwidth]{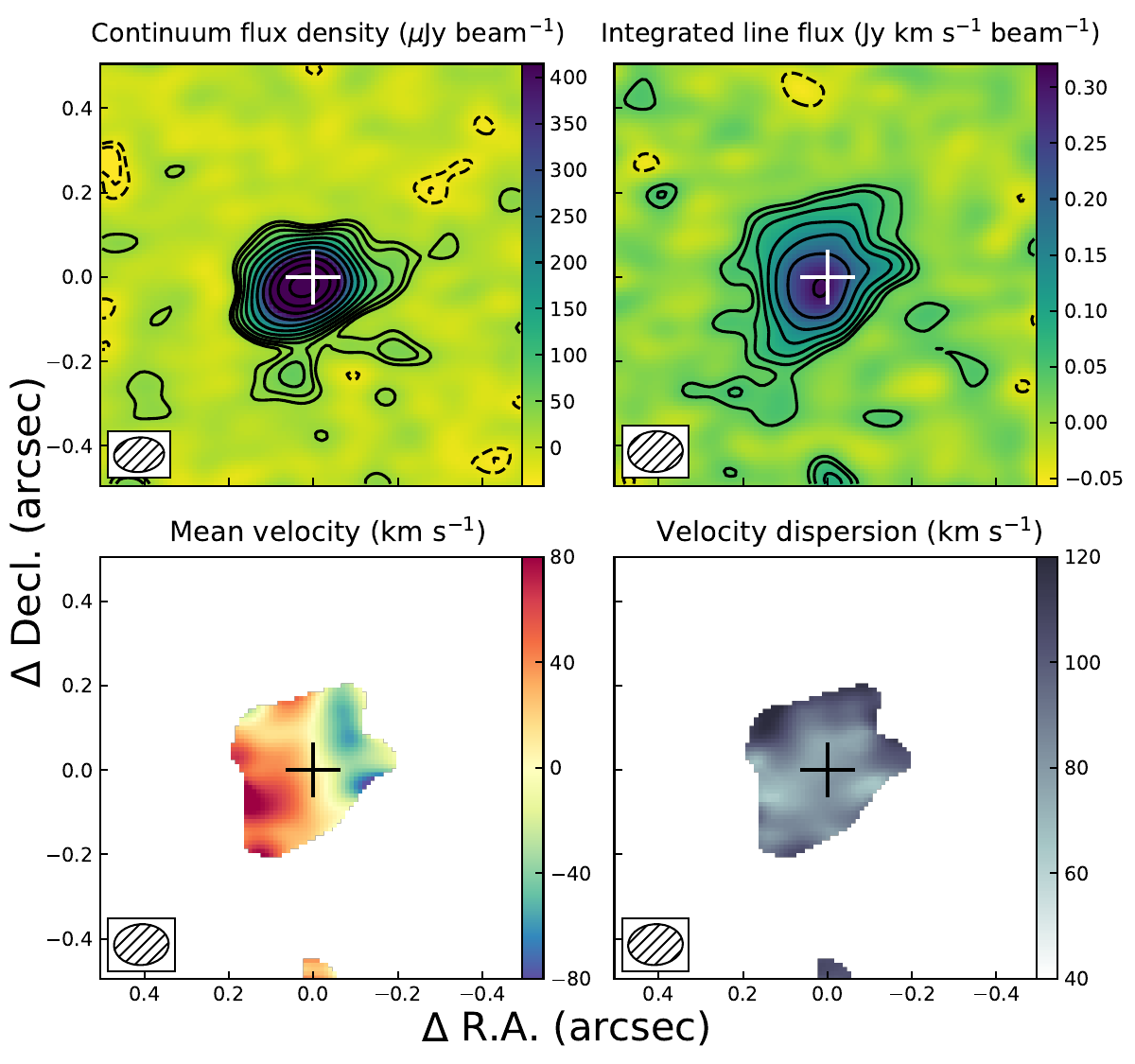}
    \caption{J2002$-$3013 continuum, \CII\ Integrated Intensity, Mean Velocity, and Velocity Dispersion maps. The cross marks the position of the quasar within the galaxy with error bars, as found by comparing JWST and GAIA data (detailed in Section \ref{subsec:astrometry}). The contours in the top two panels start at 2 $\sigma$ and increase by factors of $\sqrt{2}$. The synthesized beam is shown in the bottom left corner of each panel.}
    \label{fig:4pan 2002}
\end{figure*}

\begin{figure*}
    \centering
    \includegraphics[width=0.96\textwidth]{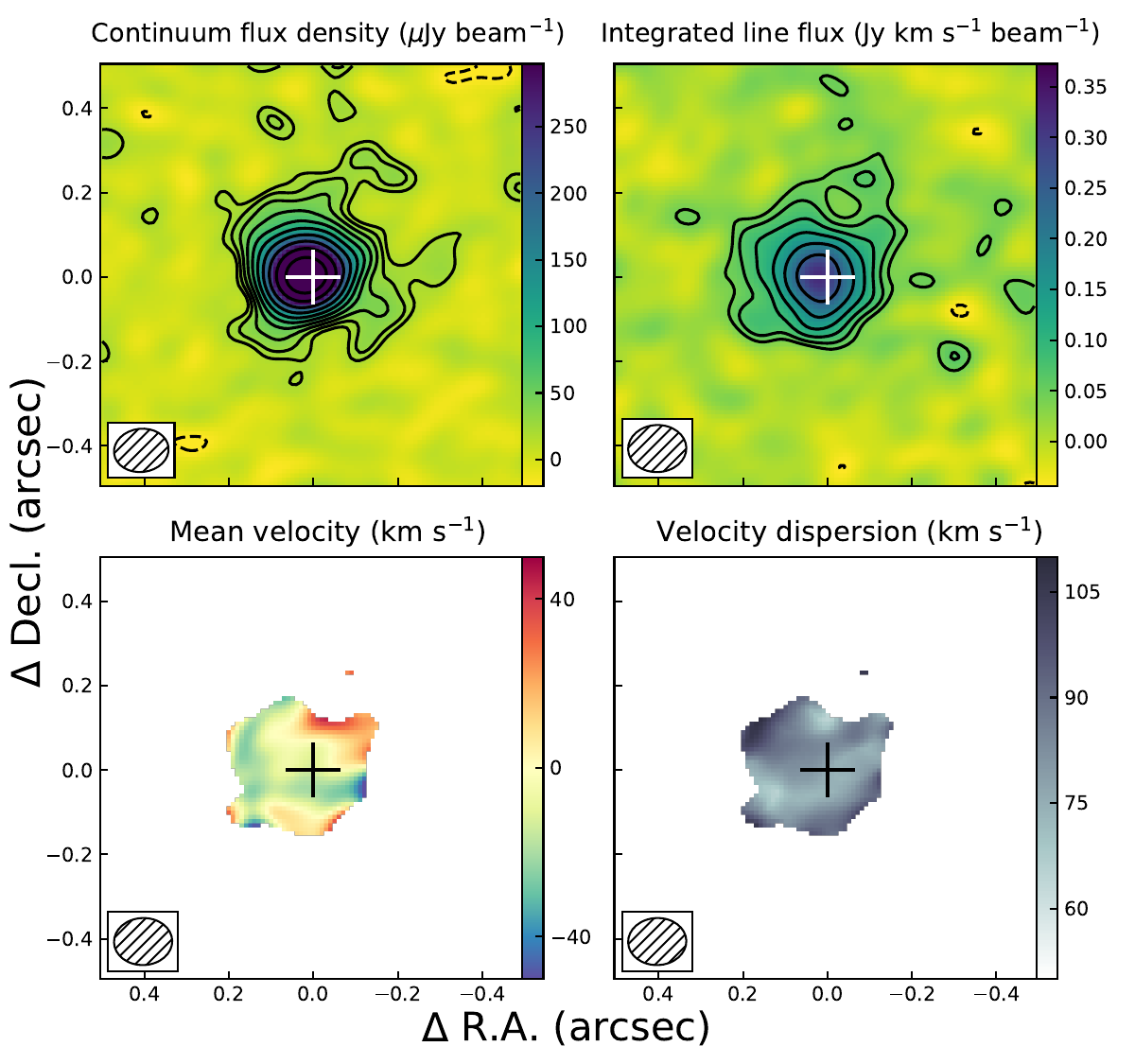}
    \caption{J2102$-$1458 continuum, \CII\ Integrated Intensity, Mean velocity, and Velocity Dispersion maps. The cross marks the position of the quasar within the galaxy with error bars, as found by comparing JWST and GAIA data (detailed in Section \ref{subsec:astrometry}). The contours in the top two panels start at 2 $\sigma$ and increase by factors of $\sqrt{2}$. The synthesized beam is shown in the bottom left corner of each panel.}
    \label{fig:4pan 2102}
\end{figure*}
\begin{figure*}
    \centering
    \includegraphics[width=0.96\textwidth]{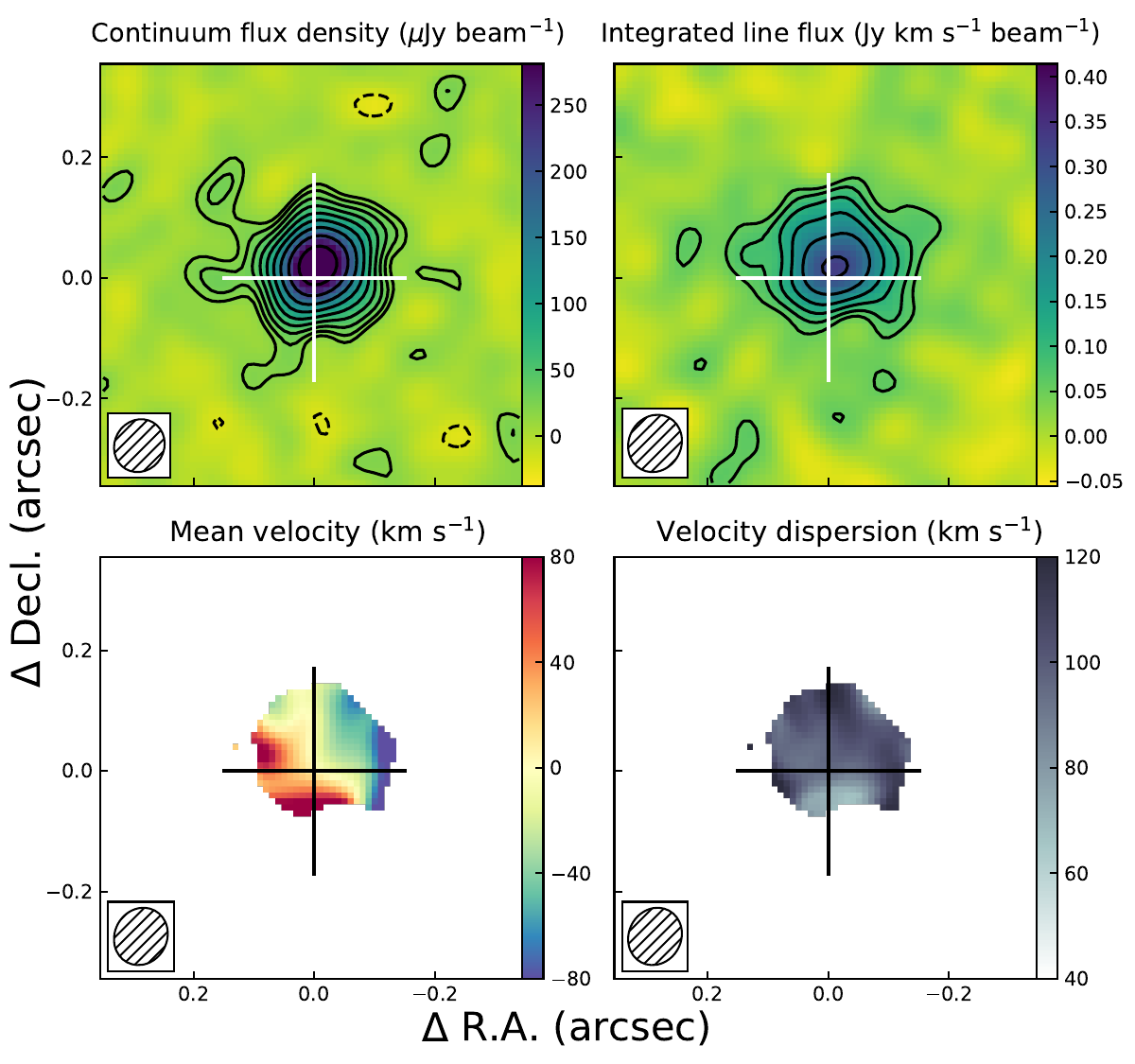}
    \caption{J0244$-$5008 continuum, \CII\ Integrated Intensity, Mean Velocity, and Velocity Dispersion maps. The cross marks the position of the quasar within the galaxy with error bars, as found by comparing JWST and GAIA data (detailed in Section \ref{subsec:astrometry}). The contours in the top two panels start at 2 $\sigma$ and increase by factors of $\sqrt{2}$. The synthesized beam is shown in the bottom left corner of each panel.}
    \label{fig: 4pan 0244}
\end{figure*}

\begin{figure*}
    \centering
    \includegraphics[width=1.0\textwidth]{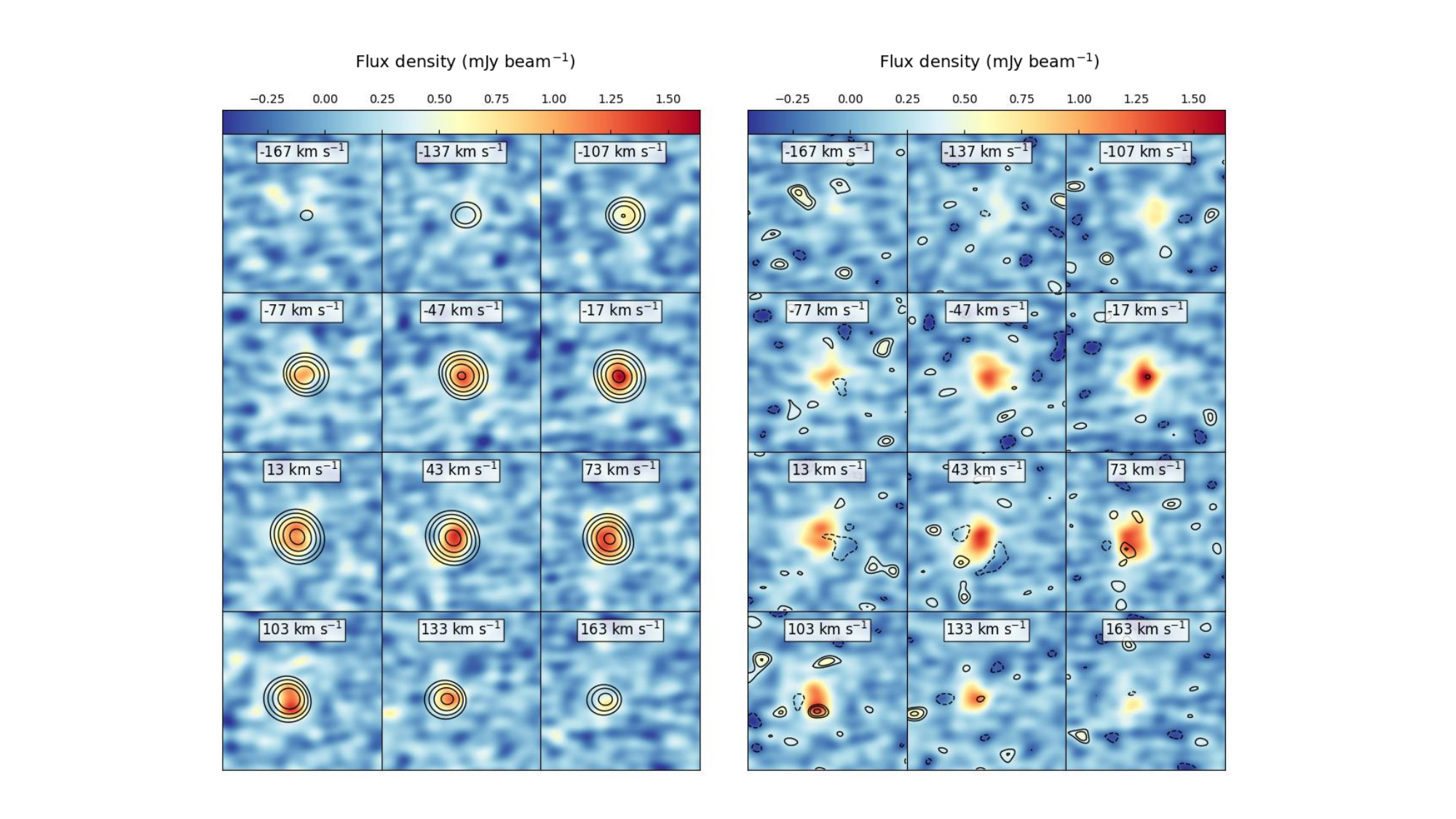}
    \caption{J2002$-$3013 Channel Maps from the \textit{qubefit} thin disk model. The left panel shows \CII\ data at each 30 \kms\ increment with contours of the model. The right panel shows \CII\ emission at each 30 \kms\ increment with residual contours. Both sets of contours start at $2 \sigma$ and increase by factors of $\sqrt{2}$.}
    \label{fig:chanmap 2002}
\end{figure*}
\begin{figure*}
    \includegraphics[width=1.1\textwidth]{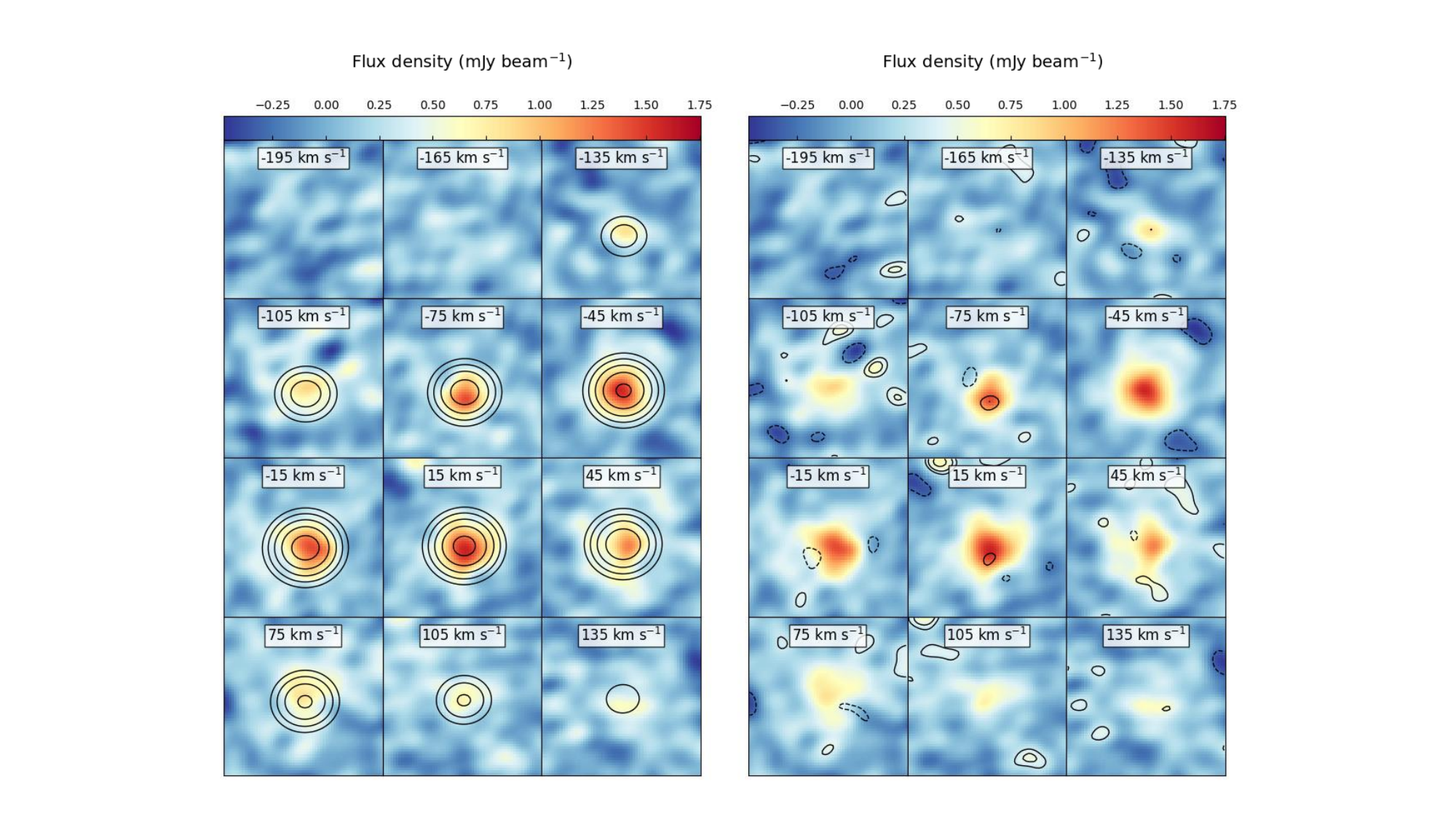}
    \caption{J2102$-$4711 Channel Maps from the \textit{qubefit} thin disk model. The left panel shows \CII\ data at each 30 \kms\ increment with contours of the model. The right panel shows \CII\ emission at each 30 \kms\ increment with residual contours. Both sets of contours start at $2 \sigma$ and increase by factors of $\sqrt{2}$.}
    \label{fig: chanmap 2102}
\end{figure*}
\begin{figure*}
    \includegraphics[width=1.1\textwidth]{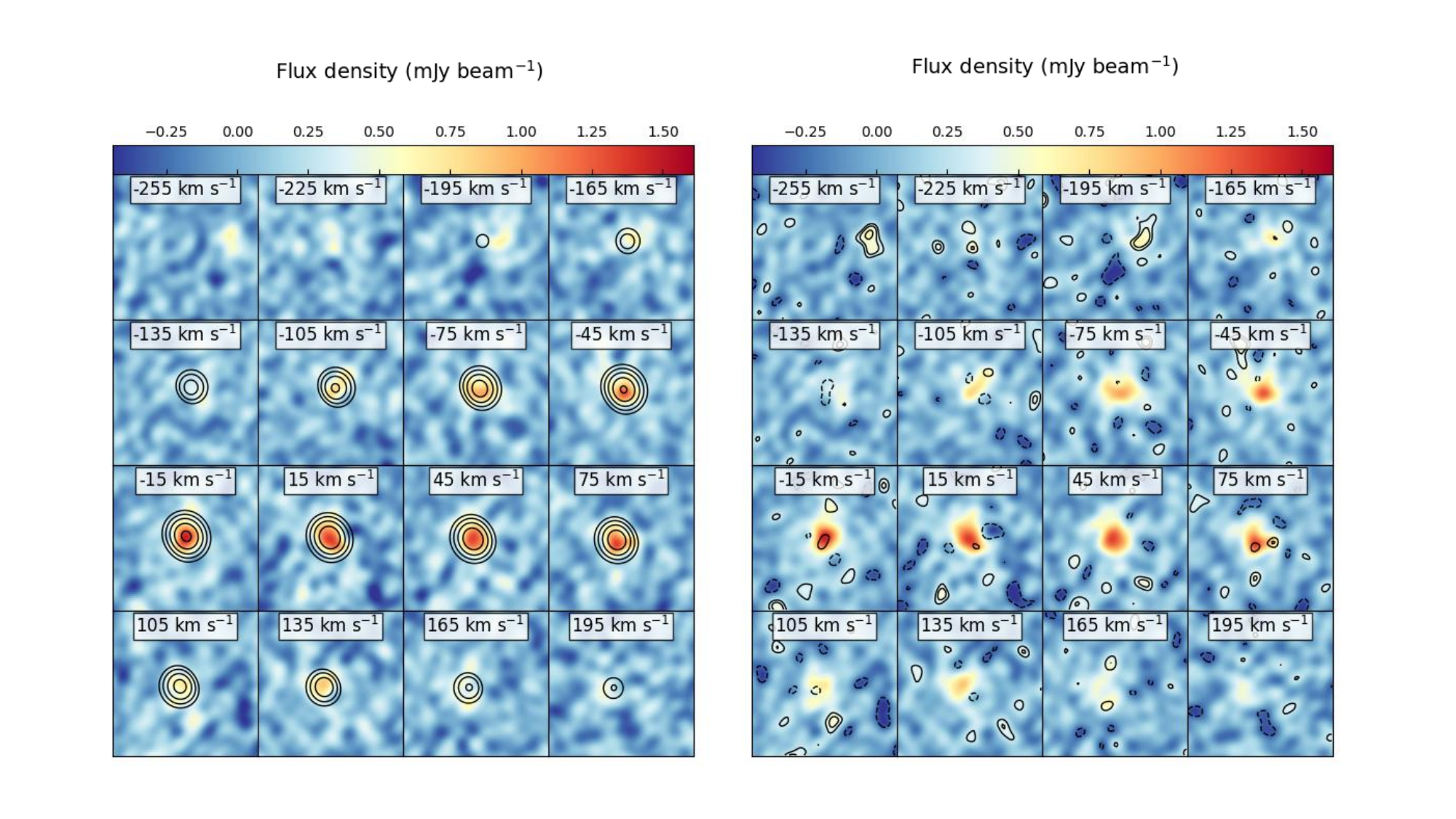}
    \caption{J0244$-$5008 Channel Maps from the \textit{qubefit} thin disk model. The left panel shows \CII\ data at each 30 \kms\ increment with contours of the model. The right panel shows \CII\ emission at each 30 \kms\ increment with residual contours. Both sets of contours start at $2 \sigma$ and increase by factors of $\sqrt{2}$.}
    \label{fig:chanmap 0244}
\end{figure*}



\end{document}